\documentclass[twocolumn,superscriptaddress]{revtex4-1}

\usepackage{amssymb}
\usepackage{amsmath}
\usepackage{graphicx}
\usepackage{xcolor, ulem}
\usepackage[version=3]{mhchem}
\usepackage{footnote}
\usepackage{physics}
\usepackage{bm}
\usepackage{dsfont}
\usepackage{cleveref}
\crefname{equation}{Eq.}{Eqs.}
\Crefname{equation}{Equation}{Equations}
\crefname{figure}{Fig.}{Figs.}
\Crefname{figure}{Figure}{Figures}
\crefname{section}{Sect.}{Sects.}
\Crefname{section}{Section}{Sections}
\Crefname{table}{Table}{Tables}

\usepackage[cal=boondox]{mathalfa}
\newcommand{\mc}{\mathcal{c}}
\newcommand{\ml}{\mathcal{l}}
\newcommand{\Cnw}{C_{\text{nw}}}
\newcommand{\Lnw}{L_{\text{nw}}}
\newcommand{\xsf}{\mathsf{x}}

\newcommand{\intdx}{\int_{-1}^{1}d\mathsf{x}}

\begin{document}
\textheight = 59\baselineskip

\title{Nanowire-Superinductance Fluxonium Qubit}

\author{T. M. Hazard}\thanks{These authors contributed equally to this work.}
\affiliation{Department of Electrical Engineering, Princeton University, Princeton, New Jersey 08544}
\author{A. Gyenis}\thanks{These authors contributed equally to this work.}
\affiliation{Department of Electrical Engineering, Princeton University, Princeton, New Jersey 08544}
\author{A. Di Paolo}\thanks{These authors contributed equally to this work.}
\affiliation{Institut quantique and D\'epartement de Physique, Universit\'e de Sherbrooke, Sherbrooke J1K 2R1 Qubec, Canada}
\author{A. T. Asfaw}
\affiliation{Department of Electrical Engineering, Princeton University, Princeton, New Jersey 08544}
\author{S. A. Lyon}
\affiliation{Department of Electrical Engineering, Princeton University, Princeton, New Jersey 08544}
\author{A. Blais}
\affiliation{Institut quantique and D\'epartement de Physique, Universit\'e de Sherbrooke, Sherbrooke J1K 2R1 Qubec, Canada}
\affiliation{Canadian Institute for Advanced Research, Toronto, Ontario, Canada}
\author{A. A. Houck}
\affiliation{Department of Electrical Engineering, Princeton University, Princeton, New Jersey 08544}

\date{\today}

\begin{abstract}
We characterize a fluxonium qubit consisting of a Josephson junction inductively shunted with a NbTiN nanowire superinductance. We explain the measured energy spectrum by means of a multimode theory accounting for the distributed nature of the superinductance and the effect of the circuit nonlinearity to all orders in the Josephson potential. Using multiphoton Raman spectroscopy, we address multiple fluxonium transitions, observe multilevel Autler-Townes splitting and measure an excited state lifetime of $T_\mathrm{1}=20$ $\mu$s.  By measuring $T_1$ at different magnetic flux values, we find a crossover in the lifetime limiting mechanism from capacitive to inductive losses.
\end{abstract}

\maketitle
The development of superinductors \cite{FluxoniumScience2009, MaslukJJarrays, PopFluxoniumLossNature, VoolFluxoniumJumps2014,BellTunableSuperInductanceLoops} has received significant interest due to their potential to provide noise protection in superconducting qubits \cite{IoffeTopologicalQubits2002,ZeroPiBrooksKiatevPreskill,KitaevZeroPi}. Moreover, inductively shunted Josephson junction based superconducting circuits are known to be immune to charge noise \cite{FluxoniumScience2009}, and to flux noise in the limit of large inductances \cite{HeavyFluxonium2017, FluxoniumForbiddenTransitionDriving, DempsterZeroPi, PeterZeroPi2017}.  Despite remarkable progress, the superinductances that have been so far reported in the literature are still small compared to those needed for qubit protection \cite{ZeroPiBrooksKiatevPreskill,KitaevZeroPi,DempsterZeroPi,PeterZeroPi2017}.

A thin-film nanowire built from a disordered superconductor constitutes an alternative approach to reach the required superinductance regime.  High-kinetic inductance superconducting materials, such as NbTiN and TiN, have been studied in the context of microwave detectors  \cite{KermanNbNdetectors2007,tunableSuperInductance2010,BerggrenNbNresonators2016}, parametric amplifiers \cite{KineticInductanceParaamp2012,NbTiNparaAmp2016,KITWPA2017}, and rfSQUID qubits \cite{KermanNanoWireQubitPRL2010,rFSquidKineticIndcutance2017}.  In a nanowire, the inertia of the Cooper pair condensate is manifested as the kinetic inductance of the superconducting wire, and can be expressed as 
\begin{equation}
L_k = \left(\frac{m}{2e^2n_s}\right)\left(\frac{l}{wd}\right),
\label{eq:kinetic_inductance}
\end{equation}
where $m$ is the free electron mass, $e$ is the electron charge and $n_s$ is the density of Cooper pairs \cite{Tinkham,tunableSuperInductance2010}.  The second bracketed term in \cref{eq:kinetic_inductance} is a geometric factor dependent on the length $l$, width $w,$ and thickness $d$ of the nanowire.  By choosing a disordered superconductor with a low $n_s$ and fabricating a sufficiently long and thin wire, the kinetic inductance can be made large enough to reach the superinductance regime.  In this regime, the presence of stray ground capacitance and the large kinetic inductance lower the frequencies of the self-resonant modes of the device. As is the case of long junction arrays \cite{MaslukJJarrays}, the multimode structure of the device needs to be taken into account to produce an accurate theoretical description \cite{FluxoniumCollectiveModes2015,JensCollectiveExcitations2013}.

In this Letter, we demonstrate a fluxonium circuit integrating a NbTiN nanowire superinductance. We characterize the effect of the nanowire modes on the qubit spectrum with a multimode circuit theory accounting for the distributed nature of the superinductance. Importantly, and in contrast to previous approaches tailored to weakly anharmonic qubits \cite{BBQPRL2012,solgun2014blackbox}, our theory incorporates the circuit nonlinearity to all orders in the Josephson potential. Such difference allows us to treat the strong anharmonicity of the fluxonium qubit efficiently, and to retain the effect of charge dispersion in the multimode Hamiltonian.

\begin{figure} [h]
\includegraphics[scale=0.338]{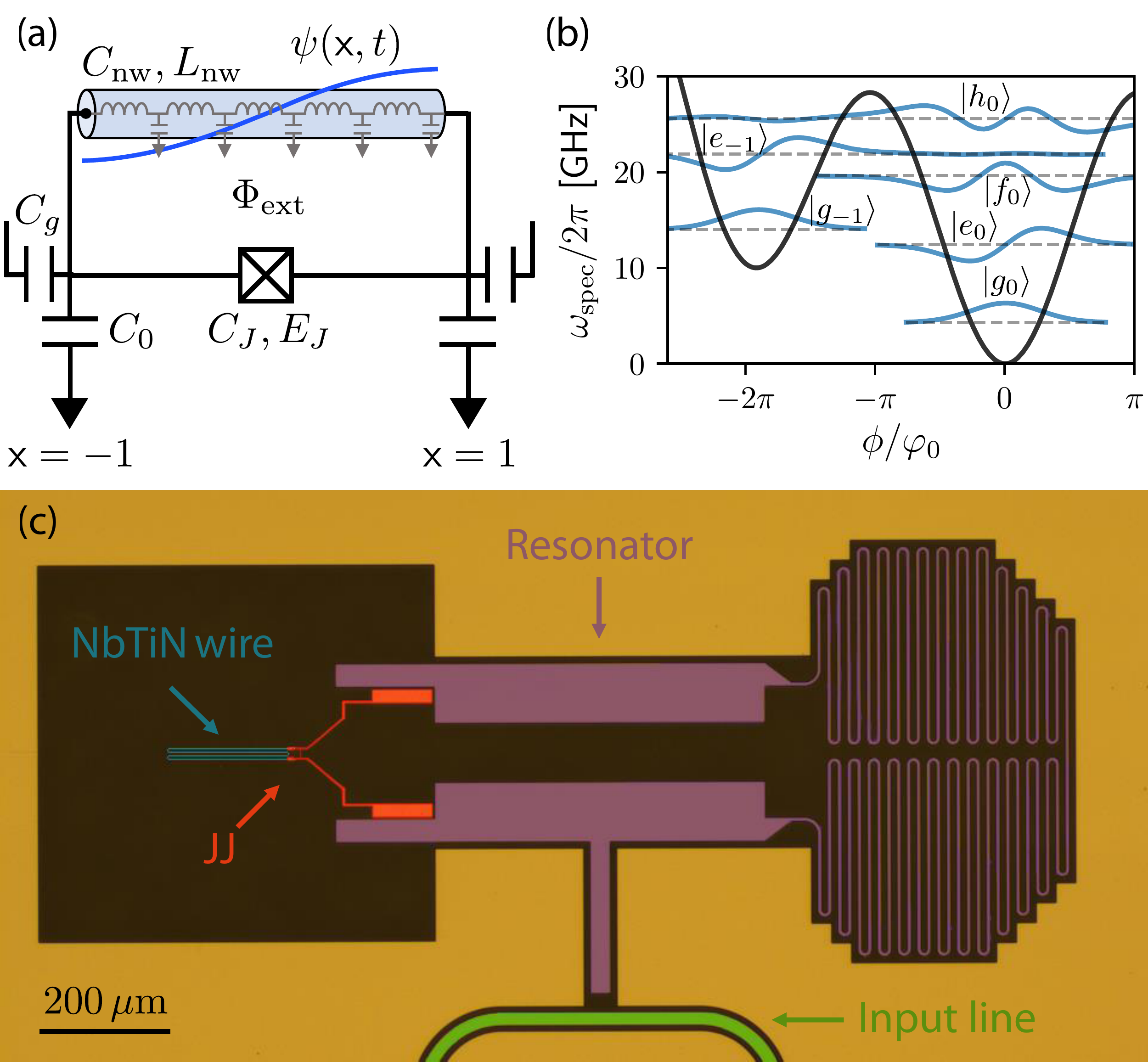}
\caption{\label{fig:fig1}(a)  The circuit diagram for the qubit, with the first antisymmetric standing wave nanowire mode in blue. $\psi(\mathsf{x},t)$ denotes the flux operator as a function of the dimensionless coordinate $\mathsf{x}=x/l$.  An off-chip coil generates the magnetic flux ($\Phi_{\text{ext}}$) that is threaded through the loop formed by the nanowire and the junction. $C_g$ and $C_0$ are the coupling capacitances to the readout resonator and to ground, respectively. (b)  The first few fluxonium eigenstates plotted for $\Phi_\mathrm{ext}/\varphi_0=-0.38 \pi$, and the respective qubit potential with wells around $\phi/\varphi_0=-2\pi$ and $\phi/\varphi_0=0$, where $\varphi_0=\hbar/2e$. (c)  False colored image of the device with the NbTiN nanowire shown in blue, the single Josephson junction and gate capacitors in red, the readout resonator in purple and the input transmission line in green.  
}
\end{figure}

A simplified circuit schematic of the nanowire superinductance fluxonium is shown in \cref{fig:fig1} (a).  In contrast to standard fluxonium devices, where a lumped element inductor shunts the Josephson junction \cite{FluxoniumScience2009,PopFluxoniumLossNature,VoolFluxoniumJumps2014,ChargingInFluxonium2009,QCPSFluxoniumValdimir2012,JensDisperssiveFluxonium2013}, our circuit model takes into account the fact that the nanowire superinductor is a high-impedance transmission line.  We present data from measurements of three devices fabricated on two different films. The nanowires in devices 1 and 2 have widths of 110 and 40 nm, respectively, equal lengths of $730\,\mu\text{m}$ and a film thickness of 15 nm.  The nanowire in device 3 is fabricated on a 10 nm thick film, has a width of 100 nm and length of $630\,\mu\text{m}$.  All the nanowires are fabricated by etching a wire pattern into the NbTiN film, with a single Al/AlO$_\text{x}$/Al junction connecting the two ends of the superinductor together. The qubit on devices 1 and 2 is capacitively coupled to a lumped element Nb resonator, with resonance frequency $\omega_r/2\pi = 6.08$ GHz and a loaded quality factor of $Q=8,400$. The qubit on device 3 is coupled to a half-wavelength coplanar waveguide resonator with $Q=14,800$ and $\omega_r/2\pi = 7.50$ GHz.  An optical image of device 1 is shown in \cref{fig:fig1} (c).

\begin{figure}[t]
\includegraphics[scale=1.065]{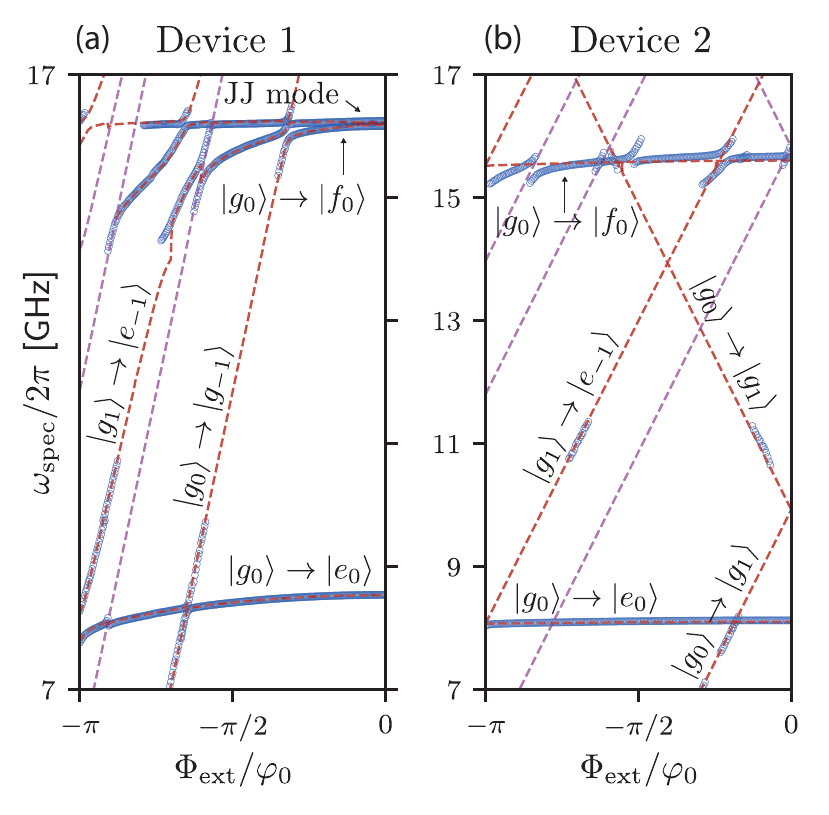}
\caption{\label{fig:fig21} Two-tone spectroscopy of device 1 (a) and device 2 (b) as a function of $\Phi_{\text{ext}}$.  The experimentally measured transition frequencies are indicated with blue markers. The result of a fit to the two-mode Hamiltonian in \cref{eq:two_mode_Hamiltonian_main} and detailed in Ref. \cite{Supplement}, is shown with red dashed lines corresponding to the fluxonium spectrum and with purple dashed lines indicating sideband transitions \cite{WallraffSideBandTransitions2007}. In (a), the inscription ``JJ mode'' (Josephson junction mode) identifies the second antisymmetric nanowire mode.}
\end{figure}

The fluxonium energy spectrum is obtained by performing two-tone spectroscopy measurements as a function of the external magnetic flux, $\Phi_{\text{ext}}$.  The amplitude of the transmitted power is monitored at the dressed cavity frequency while sweeping a second spectroscopic tone of frequency $\omega_\mathrm{spec}/2\pi$.  The measurement results are shown in \cref{fig:fig21}.  Labeling the energy eigenstates within a single potential well as $\ket{g_i},\ket{e_i},\ket{f_i},...$, where the index $i$ indicates the potential well to which these belong [see \cref{fig:fig1} (b)], the fluxonium transitions are classified in two types:  intrawell plasmons, such as $\ket{g_0}\rightarrow \ket{e_0},$ and interwell fluxons, such as $\ket{g_0}\rightarrow \ket{g_{-1}}$.  Parity selection rules of the fluxonium circuit allow for transitions between adjacent plasmon states by absorption of a single photon.  However, the direct transition $\ket{g_0}\rightarrow \ket{f_0}$ can only be completed via a two-photon process in which $\ket{e_0}$ serves as an intermediate virtual state.  We note that devices 1 and 2 operate in a similar parameter regime to ``heavy fluxonium" \cite{HeavyFluxonium2017,FluxoniumSelectrionRuleEngineering2018}, where the ratio between the Josephson ($E_J$) and charging ($E_C$) energies is large.  As a consequence, transitions between the fluxonium potential wells are exponentially attenuated.  Therefore, such excitations are most clearly visible in the regions where they hybridize with the plasmon energy levels.

\cref{fig:fig21} (a) shows the presence of a second fluxonium mode for device 1 at 16.3 GHz. While similar characteristics have been observed in previous fluxonium devices, high-frequency modes have been so far phenomenologically modeled as harmonic oscillators linearly coupled to the qubit degree of freedom \cite{FluxoniumScience2009}. Here we go beyond such an approximation and derive a multimode Hamiltonian considering the complete device Lagrangian, which accounts for the distributed nature of the superinductance. Importantly, we find that the qubit spectrum is determined by the nonlinear interaction of the circuit modes which are antisymmetric at the Josephson junction ports [see \cref{fig:fig1} (a)].  The agreement with the measured data is excellent over a very large frequency range. 

The nanowire is described as a homogeneous transmission line with distributed capacitance $\mc=\Cnw/2l$ and inductance $\ml=\Lnw/2l,$ where $\Cnw,$ $\Lnw$, and $2l$ are, respectively, the total ground capacitance, inductance, and length of the nanowire. Defining the flux operator $\psi(\xsf,t)$ in terms of the dimensionless coordinate $\xsf=x/l$, the nanowire Lagrangian can be written as
\begin{equation}
\mathcal{L}_{\text{nw}} = \intdx \frac{(\Cnw/2)}{2}\dot{\psi}(\xsf,t)^2 - \frac{1}{2(\Lnw/2)}{\psi}(\xsf,t)^2.
\label{eq:nw_field_Lagrangian}
\end{equation}
Additionally, we consider gate capacitances ($C_g$) placed at the two ports of the device ($\xsf_p=\pm 1$) with respective driving voltages $\{V_{\xsf_p}\}$, as well as ground capacitances ($C_0$). The Lagrangian of the inductively shunted Josephson junction then reads 
\begin{equation}
\begin{split}
\mathcal{L} = & \sum_{\xsf_p} \frac{C_g}{2}\qty(\dot{\psi}(\xsf_p,t)-V_{\xsf_p})^2 + \frac{C_0}{2} \dot{\psi}(\xsf_p,t)^2\\
&+\mathcal{L}_{\text{nw}} + \frac{C_J}{2} \dot{\delta}_{\psi}(t)^2 + E_J\cos({\delta}_{\psi}(t)/\varphi_0),
\end{split}
\label{eq:device_Lagrangian_main}
\end{equation}
where 
\begin{equation}
{\delta}_{\psi}(t)/\varphi_0 = (\Delta\psi(t) + \Phi_{\text{ext}})/\varphi_0 ,
\label{eq:JJ_flux_difference}
\end{equation}
is the gauge-invariant superconducting phase difference across the junction, $\Delta\psi(t) = {\psi}(1,t)-{\psi}(-1,t)$ is the flux operator difference at the boundaries of the superinductor, and $E_J$ is the Josephson energy \cite{JJembededInTXLine2012,NormalModesTXResonator2016}. 

To obtain a tractable theoretical description of our device, we map \cref{eq:device_Lagrangian_main} into the Lagrangian of an infinite number of nonlinearly interacting normal modes \cite{Supplement}. We observe that modes which are symmetric at the junction ports are not coupled to the Josephson nonlinearity, and thus do not contribute to the qubit Hamiltonian. We therefore derive a multimode Hamiltonian for the antisymmetric normal modes, which is later truncated to a finite number of modes. The truncation is possible due to the fact that only few antisymmetric modes lie in the frequency range of interest. Furthermore, the effective normal mode impedance decreases quickly with the mode number such that high-frequency modes are only weakly anharmonic. 

We find that the spectra of our devices can be accurately described by a two-mode Hamiltonian of the form
\begin{equation}
\begin{split}
\hspace*{-0.07cm}H_{\text{two-mode}} =& \frac{(q_{0}-q_{g0})^2}{2\tilde{C}_{0}} + \frac{\phi_{0}^2}{2\tilde{L}_{0}} + \frac{(q_{1}-q_{g1})^2}{2\tilde{C}_{1}} + \frac{\phi_{1}^2}{2\tilde{L}_{1}}\\
&-\frac{\phi_{0}\phi_{1}}{L_J} - E_J \cos(\frac{\phi_{0}+\phi_{1}}{\varphi_0}+\frac{\Phi_{\text{ext}}}{\varphi_0}),
\end{split}
\label{eq:two_mode_Hamiltonian_main}
\end{equation}
\noindent where $\tilde{C}_i,$ $\tilde{L}_i$, and $q_{gi}$ are, respectively, the effective capacitance, inductance and offset charge corresponding to the first two antisymmetric modes labeled by $i=\{0,1\}$ and $L_J=E_J/\varphi_0^2$. The definitions of the various parameters in \cref{eq:two_mode_Hamiltonian_main} is provided in Ref.~\cite{Supplement}. The results in \cref{fig:fig21} are obtained by numerical diagonalization of the complete Hamiltonian of the device, including \cref{eq:two_mode_Hamiltonian_main}, the resonator Hamiltonian and the interaction between such systems \cite{Supplement}. 

From our two-mode fit to the qubit spectrum, we find nanowire inductances of 121 nH, 314 nH and 309 nH for devices 1, 2, and 3, respectively, and corresponding characteristic impedances ($Z_{\text{nw}}=\sqrt{L_{\text{nw}}/C_{\text{nw}}}$) of about 1.85, 7.38, and 12.43 k$\Omega$. The inductance values from the fit are within 7 \% of the theoretical prediction given by \cref{eq:kinetic_inductance} \cite{Supplement}.  \Cref{tab:devParams} provides the Hamiltonian parameters extracted from a single-mode fit allowing direct comparison to previous implementations of JJ array based fluxonium devices \cite{FluxoniumScience2009,FluxoniumForbiddenTransitionDriving,FluxoniumSelectrionRuleEngineering2018,HeavyFluxonium2017,PopFluxoniumLossNature}.
\begin{center}
\begin{table}[h]
\begin{tabular}{|c|c|c|c|} 
\hline
Device & $E_C$ [GHz] & $E_L$ [GHz] & $E_J$ [GHz] \\
\hline\hline
1 & 0.89 & 1.37 & 10.95 \\ 
\hline
2 & 0.56 & 0.52 & 16.16 \\
\hline
3 & 1.90 & 0.53 &  5.90 \\
\hline
\end{tabular}
\caption{\label{tab:devParams} Device parameter table obtained from a single-mode fit to the fluxonium qubit spectrum, for devices 1, 2, and 3.} 
\end{table}
\end{center}

\begin{figure}[h]
\includegraphics[scale=0.26]{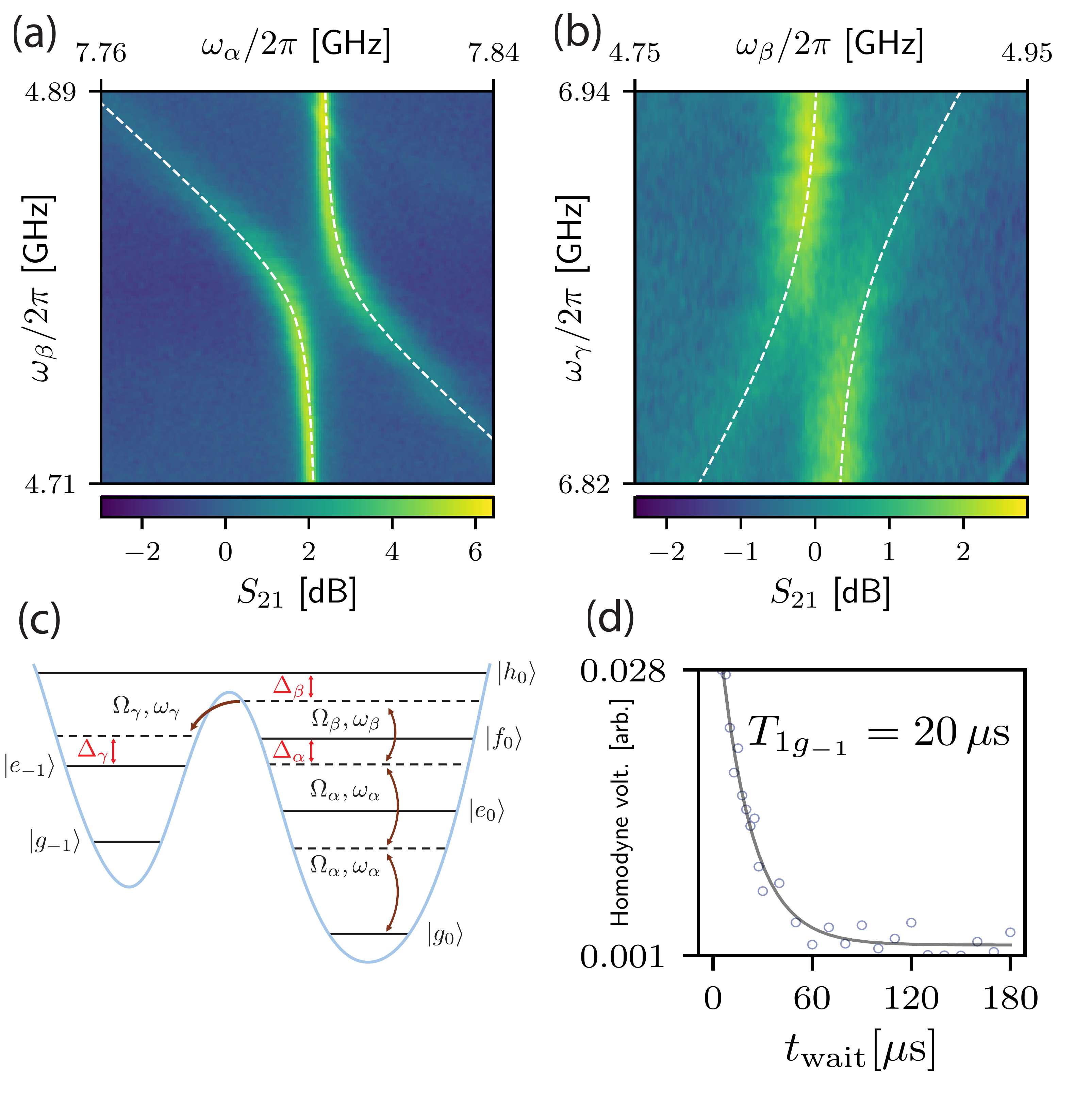}
\caption{\label{fig:autlerTownes} The multitone spectroscopy data, taken at $\Phi_{\text{ext}}/\varphi_0=-0.46\pi$, demonstrating population transfer between $\ket{g_0}$ and $\ket{h_0}$ (a) with $\Omega_\gamma=0$,  and $\ket{h_0}$ to $\ket{e_{-1}}$ (b) with fixed $\omega_\alpha/2\pi=7.78$ GHz. The white dashed lines indicate the maximum population from a multilevel master equation simulation \cite{Supplement}.  (c) A schematic diagram of the device 2 level structure in the presence of coherent external drives.  The drives, with frequencies $\omega_i/2\pi$ and amplitudes $\Omega_i$ are detuned from the levels by $\Delta_i/2\pi$.
(d)  Three sequential $\pi$ pulses ($\sigma=15$ ns) are applied at the transition frequencies to perform $T_1$ measurements of the $\ket{g_{-1}}$ state.  The demodulated homodyne voltage from the readout resonator is measured as a function of $t_\mathrm{wait}$.}
\end{figure}

In devices 1 and 2, the small dipole element between the fluxon states makes it experimentally challenging to directly drive the $\ket{g_{-1}}\rightarrow\ket{g_{0}}$ transition.  By using multiple drives, we are able to transfer the ground state population between the neighboring wells using the intermediate $\ket{h_0}$ state, which is located close to the top of the barrier and has spectral weight in both wells.  We apply three coherent and simultaneous drives of frequencies $\omega_\alpha/2\pi$, $\omega_\beta/2\pi$ and $\omega_\gamma/2\pi$, respectively, targeting the $\ket{g_0}\rightarrow \ket{f_0}$ (two-photon), the $\ket{f_0}\rightarrow \ket{h_0}$ (one-photon), and the $\ket{h_0}\rightarrow \ket{e_{-1}}$ (one-photon) transitions [see \cref{fig:autlerTownes}(a)].

At $\Phi_{\text{ext}}/\varphi_0=-0.46\pi$, we set $\Omega_\gamma=0$ and simultaneously vary $\omega_\alpha/2\pi$ and $\omega_\beta/2\pi$ around the $\ket{g_0}\rightarrow \ket{f_0}$ and $\ket{f_0}\rightarrow \ket{h_0}$ transitions.  We observe a vertical band corresponding to the $\ket{g_0}\rightarrow \ket{f_0}$ transition at 7.8 GHz, and a diagonal band with a slope of $\omega_\alpha/\omega_\beta=-1/2,$ corresponding to the Raman transition between the $\ket{g_0}$ and $\ket{h_0}$ states [\cref{fig:autlerTownes} (a)].  Around the resonance condition ($2\hbar\omega_\alpha\approx E_{f_0}-E_{g_0}$ and $\hbar\omega_\beta\approx E_{h_0}-E_{f_0}$), the two bands exhibit an avoided crossing, which is the hallmark of the Autler-Townes doublet previously observed in other superconducting qubits \cite{WallraffATsplitting2009,HakonenAT3levels2009,AT3DtransmonPalmer,PalmerATNJP2013}. Next, we fix the frequency of the $\alpha$ tone at $\Delta_\alpha/2\pi=20$ MHz, turn on the $\gamma$ drive, and simultaneously scan the frequencies $\omega_\beta/2\pi$ and $\omega_\gamma/2\pi$. \Cref{fig:autlerTownes} (b) displays the resulting Autler-Townes splitting, where the Raman transition manifests itself here with a slope of $\omega_\gamma/\omega_\beta=+1,$ corresponding to the three-drive Raman condition.  This method allows us to experimentally determine the energy levels of the fluxonium qubit using population transfer.

With complete information regarding the energy of the fluxonium excited states, we determine the relaxation rate of the $\ket{g_{-1}}$ state by performing time-resolved measurements \cite{MultilevelPiPulse2015}.  We use the frequency values obtained from the Raman spectroscopy and perform a pulse sequence which consists of three sequential $\pi$ pulses at the transition frequencies $\left(E_{f_0}-E_{g_0}\right)/h$, $\left(E_{h_0}-E_{f_0}\right)/h$ and $\left(E_{h_0}-E_{e_{-1}}\right)/h$ to prepare the system in the $\ket{e_{-1}}$ state.  At the end of this procedure, the system relaxes into the $\ket{g_{-1}}$ state, on the timescale of the plasmon $T_1$ ($\sim600$ ns).  On a longer timescale, the system relaxes back to $\ket{g_0}$.  For $t_\mathrm{wait}\gg T_\mathrm{1e_{0}}$, the reduction in $\ket{g_{-1}}$ population follows an exponential decay with $T_\mathrm{1g_{-1}}=$ 20 $\mu$s.

\begin{figure}[t!]
\includegraphics[scale=0.355]{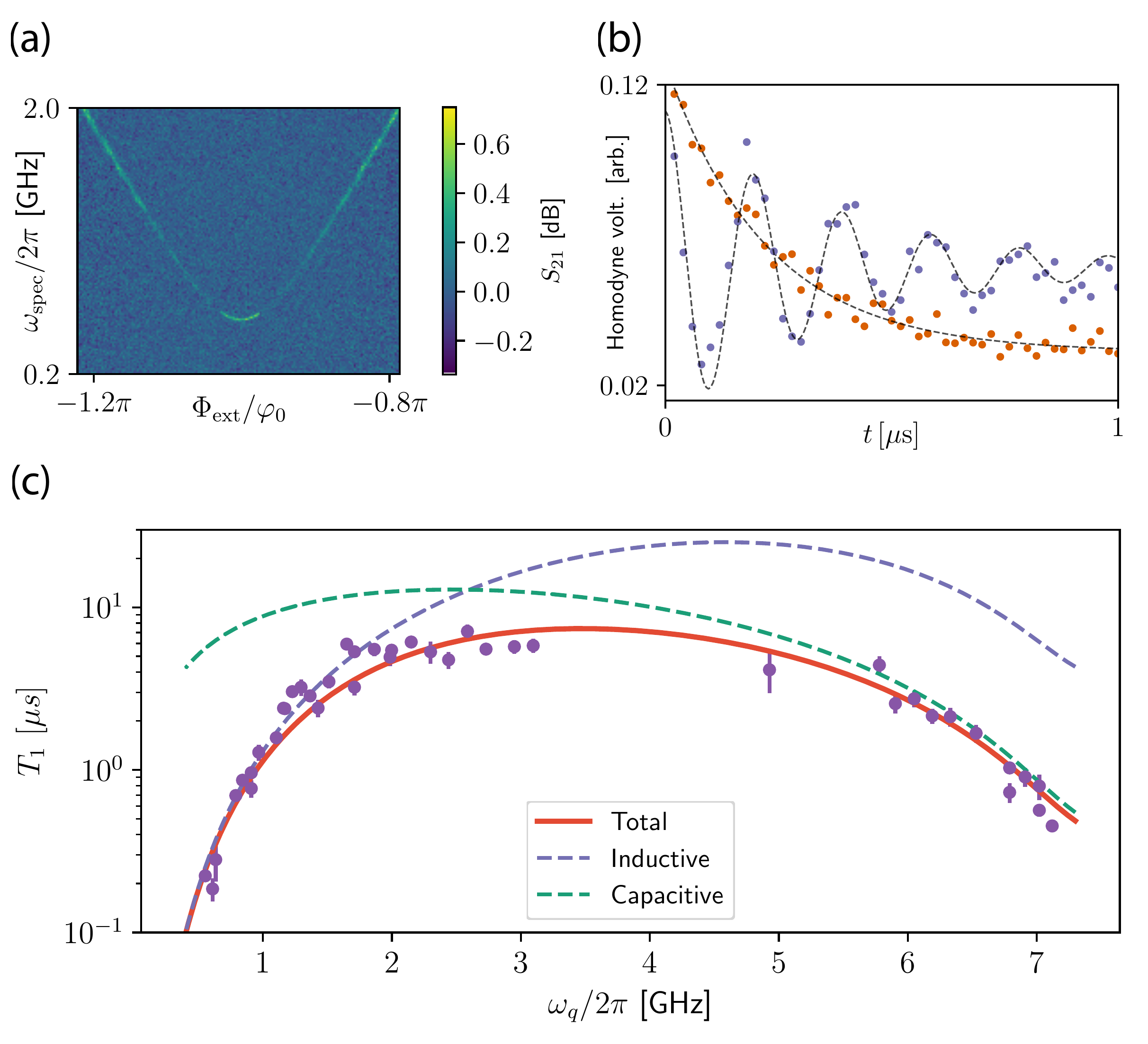}
\caption{\label{fig:fig5} (a)  Low frequency spectroscopy data from device 3.  (b)  $T_1$ (red) and $T_\mathrm{2Ramsey}$ (blue)  data taken at $\Phi_{\text{ext}}/\varphi_0=-\pi$.  (c)  $T_1$ as a function of qubit frequency.  The lines represent the theory fits for total (red), inductive (blue), and capacitive (green) $T_1$.  The $T_1$ values were obtained with both pulsed and mixed state driving.  Measurements using both types of excited state preparation at the same flux gave the same value of $T_1$.}
\end{figure}

Because of the high $E_J/E_C$ ratio, devices 1 and 2 lack flux insensitive sweet spots at zero and half flux.  In order to fully characterize the coherence properties of the qubit and demonstrate coherent control between the fluxon states, we reduced the $E_J/E_C$ ratio in device 3. The overlap between the fluxon wave functions is made sufficiently large to directly observe the transition with a one-photon drive, which comes at the cost of increased sensitivity to different relaxation mechanisms.  The low frequency, two-tone spectroscopy data for device 3 are shown in \cref{fig:fig5}.  At $\Phi_{\text{ext}}/\varphi_0=-\pi$, the spectrum shows a flux-insensitive fluxon transition, where we perform coherence measurements and find $T_1=220$ ns, $T_\mathrm{2Ramsey}=380$ ns and $T_\mathrm{2Echo}\approx2T_1$ indicating that the qubit dephasing is dominated by qubit relaxation.

By changing $\Phi_{\text{ext}}$, we measure $T_1$ of the fluxon transition as a function of qubit frequency. The data show an increase in $T_1$ as the qubit frequency is increased to a maximal value of 7 $\mu$s for frequencies between 2-3 GHz.  Upon further increasing the qubit frequency, $T_1$ decreases by an order of magnitude [\cref{fig:fig5}c].

To understand the $T_1$ frequency dependence, we take into account inductive and capacitive loss mechanisms, which can be described with the following expressions:

\begin{equation}
\Gamma_\mathrm{ind}=\frac{E_L}{\hbar Q_L}\left(\coth\left(\frac{\hbar\omega_q}{2k_BT}\right)+1\right)|\bra{g_{-1}}\hat{\varphi}\ket{g_{0}}|^2
\label{eq:IndLoss}
\end{equation} 
\begin{equation}
\Gamma_\mathrm{cap}=\frac{\hbar\omega_q^2}{8E_CQ_C}\left(\coth\left(\frac{\hbar\omega_q}{2k_BT}\right)+1\right)|\bra{g_{-1}}\hat{\varphi}\ket{g_{0}}|^2
\label{eq:CapLoss}
\end{equation} 
where $|\bra{g_{-1}}\hat{\varphi}\ket{g_{0}}|^2$ is the transition matrix element between the fluxon states, $Q_L$ and $Q_C$ are the inductive and capacitive quality factors, respectively, $k_B$ is the Boltzmann constant, $T$ is the temperature and $\omega_q$ is the fluxon transition frequency \cite{SchoelkopfQubitsAsNoiseSpec2002}.  Based on previously reported measurements \cite{PopFluxoniumLossNature}, the lifetime limitation from nonequilibrium quasiparticles is at least an order of magnitude larger than the observed relaxation times at all frequencies and is therefore not considered.  Radiative loss due to the Purcell effect \cite{AndrewSpontaneousEmission} is only significant when the qubit frequency is within $\sim$ 50 MHz of $\omega_r/2\pi= 7.5$ GHz \cite{Supplement}.   \cref{fig:fig5}c shows the measured $T_1$ (blue markers) values along with the fitted $T_1=\left(\Gamma_\mathrm{cap}^{-1}+\Gamma_\mathrm{ind}^{-1}\right)^{-1}$ (red line).  The fit of $T_1$ vs $\omega_q$ in \cref{fig:fig5}, gives $Q_L=39,000$ and $Q_C=15,100$, where the lifetime at low $\omega_q$ is dominated by inductive loss and at high $\omega_q$ by capacitive loss.  The inductor can be modeled as a lossless inductor in series with a frequency dependent resistor, where $R=\omega L/Q_\mathrm{ind}$ corresponds to $R=$ 27 m$\Omega$ at $\omega/2\pi=$ 550 MHz.  The possible sources of the inductive loss can arise from a finite contact resistance between the NbTiN wire and the Al Josephson junction leads, loss from charge impurities on the surface of the wire, or some intrinsic loss from the bulk NbTiN material.  In future devices, the geometry of the Al/NbTiN contact and nanowire dimensions could be modified to better determine what limits the inductive quality factor.  Improvements to $Q_C$ could be made by moving to a 3D architecture, where the electric field participation at lossy interfaces is reduced \cite{3DtmonYale2011}.

In conclusion, we have fabricated and measured a nanowire superinductance fluxonium qubit.  We find that the transition energy levels are modified due to the distributed nature of the nanowire, which is well explained in the framework of a multimode theory.  As the modes of the nanowire strongly depend on the parasitic and stray capacitances of the wire, using a shorter wire with higher sheet inductance (for example high quality granular aluminum films with one hundred times larger $L_k=2$ nH/$\square$ \cite{QuasiparticlesInGranularAluminum2018,GranularAlNonlinearity2018,GralFluxonium2018}), or integrating the fluxonium into a 3D cavity or waveguide \cite{3DFluxonium2017}, could reduce unwanted capacitances and help to push the nanowire self-resonant modes to higher frequencies.  The multimode theory developed here is an important step towards understanding large circuits beyond the lumped element approximation, such as the $0-\pi$ qubit \cite{ZeroPiBrooksKiatevPreskill,KitaevZeroPi}, where the distributed nature of the circuit elements is critical to device design.  

\section{Acknowledgments}
We thank Andrei Vrajitoarea, Zhaoqi Leng and J\'er\^ome Bourassa for useful discussions.  Research was supported by the Army Research Office Grant No.~W911NF-15-1-0421 and the Princeton Center for Complex Materials DMR-142052. This work was undertaken thanks in part to funding from NSERC and the Canada First Research Excellence Fund.


\begin{thebibliography}{45}%
\makeatletter
\providecommand \@ifxundefined [1]{%
 \@ifx{#1\undefined}
}%
\providecommand \@ifnum [1]{%
 \ifnum #1\expandafter \@firstoftwo
 \else \expandafter \@secondoftwo
 \fi
}%
\providecommand \@ifx [1]{%
 \ifx #1\expandafter \@firstoftwo
 \else \expandafter \@secondoftwo
 \fi
}%
\providecommand \natexlab [1]{#1}%
\providecommand \enquote  [1]{``#1''}%
\providecommand \bibnamefont  [1]{#1}%
\providecommand \bibfnamefont [1]{#1}%
\providecommand \citenamefont [1]{#1}%
\providecommand \href@noop [0]{\@secondoftwo}%
\providecommand \href [0]{\begingroup \@sanitize@url \@href}%
\providecommand \@href[1]{\@@startlink{#1}\@@href}%
\providecommand \@@href[1]{\endgroup#1\@@endlink}%
\providecommand \@sanitize@url [0]{\catcode `\\12\catcode `\$12\catcode
  `\&12\catcode `\#12\catcode `\^12\catcode `\_12\catcode `\%12\relax}%
\providecommand \@@startlink[1]{}%
\providecommand \@@endlink[0]{}%
\providecommand \url  [0]{\begingroup\@sanitize@url \@url }%
\providecommand \@url [1]{\endgroup\@href {#1}{\urlprefix }}%
\providecommand \urlprefix  [0]{URL }%
\providecommand \Eprint [0]{\href }%
\providecommand \doibase [0]{http://dx.doi.org/}%
\providecommand \selectlanguage [0]{\@gobble}%
\providecommand \bibinfo  [0]{\@secondoftwo}%
\providecommand \bibfield  [0]{\@secondoftwo}%
\providecommand \translation [1]{[#1]}%
\providecommand \BibitemOpen [0]{}%
\providecommand \bibitemStop [0]{}%
\providecommand \bibitemNoStop [0]{.\EOS\space}%
\providecommand \EOS [0]{\spacefactor3000\relax}%
\providecommand \BibitemShut  [1]{\csname bibitem#1\endcsname}%
\let\auto@bib@innerbib\@empty
\bibitem [{\citenamefont {Manucharyan}\ \emph {et~al.}(2009)\citenamefont
  {Manucharyan}, \citenamefont {Koch}, \citenamefont {Glazman},\ and\
  \citenamefont {Devoret}}]{FluxoniumScience2009}%
  \BibitemOpen
  \bibfield  {author} {\bibinfo {author} {\bibfnamefont {V.~E.}\ \bibnamefont
  {Manucharyan}}, \bibinfo {author} {\bibfnamefont {J.}~\bibnamefont {Koch}},
  \bibinfo {author} {\bibfnamefont {L.~I.}\ \bibnamefont {Glazman}}, \ and\
  \bibinfo {author} {\bibfnamefont {M.~H.}\ \bibnamefont {Devoret}},\
  }\href@noop {} {\bibfield  {journal} {\bibinfo  {journal} {Science}\ }\textbf
  {\bibinfo {volume} {326}},\ \bibinfo {pages} {113} (\bibinfo {year}
  {2009})}\BibitemShut {NoStop}%
\bibitem [{\citenamefont {Masluk}\ \emph {et~al.}(2012)\citenamefont {Masluk},
  \citenamefont {Pop}, \citenamefont {Kamal}, \citenamefont {Minev},\ and\
  \citenamefont {Devoret}}]{MaslukJJarrays}%
  \BibitemOpen
  \bibfield  {author} {\bibinfo {author} {\bibfnamefont {N.~A.}\ \bibnamefont
  {Masluk}}, \bibinfo {author} {\bibfnamefont {I.~M.}\ \bibnamefont {Pop}},
  \bibinfo {author} {\bibfnamefont {A.}~\bibnamefont {Kamal}}, \bibinfo
  {author} {\bibfnamefont {Z.~K.}\ \bibnamefont {Minev}}, \ and\ \bibinfo
  {author} {\bibfnamefont {M.~H.}\ \bibnamefont {Devoret}},\ }\href@noop {}
  {\bibfield  {journal} {\bibinfo  {journal} {Phys. Rev. Lett.}\ }\textbf
  {\bibinfo {volume} {109}},\ \bibinfo {pages} {137002} (\bibinfo {year}
  {2012})}\BibitemShut {NoStop}%
\bibitem [{\citenamefont {Pop}\ \emph {et~al.}(2014)\citenamefont {Pop},
  \citenamefont {Geerlings}, \citenamefont {Catelani}, \citenamefont
  {Schoelkopf}, \citenamefont {Glazman},\ and\ \citenamefont
  {Devoret}}]{PopFluxoniumLossNature}%
  \BibitemOpen
  \bibfield  {author} {\bibinfo {author} {\bibfnamefont {I.~M.}\ \bibnamefont
  {Pop}}, \bibinfo {author} {\bibfnamefont {K.}~\bibnamefont {Geerlings}},
  \bibinfo {author} {\bibfnamefont {G.}~\bibnamefont {Catelani}}, \bibinfo
  {author} {\bibfnamefont {R.~J.}\ \bibnamefont {Schoelkopf}}, \bibinfo
  {author} {\bibfnamefont {L.~I.}\ \bibnamefont {Glazman}}, \ and\ \bibinfo
  {author} {\bibfnamefont {M.~H.}\ \bibnamefont {Devoret}},\ }\href@noop {}
  {\bibfield  {journal} {\bibinfo  {journal} {Nature}\ }\textbf {\bibinfo
  {volume} {508}},\ \bibinfo {pages} {369} (\bibinfo {year}
  {2014})}\BibitemShut {NoStop}%
\bibitem [{\citenamefont {Vool}\ \emph {et~al.}(2014)\citenamefont {Vool},
  \citenamefont {Pop}, \citenamefont {Sliwa}, \citenamefont {Abdo},
  \citenamefont {Wang}, \citenamefont {Brecht}, \citenamefont {Gao},
  \citenamefont {Shankar}, \citenamefont {Hatridge}, \citenamefont {Catelani},
  \citenamefont {Mirrahimi}, \citenamefont {Frunzio}, \citenamefont
  {Schoelkopf}, \citenamefont {Glazman},\ and\ \citenamefont
  {Devoret}}]{VoolFluxoniumJumps2014}%
  \BibitemOpen
  \bibfield  {author} {\bibinfo {author} {\bibfnamefont {U.}~\bibnamefont
  {Vool}}, \bibinfo {author} {\bibfnamefont {I.~M.}\ \bibnamefont {Pop}},
  \bibinfo {author} {\bibfnamefont {K.}~\bibnamefont {Sliwa}}, \bibinfo
  {author} {\bibfnamefont {B.}~\bibnamefont {Abdo}}, \bibinfo {author}
  {\bibfnamefont {C.}~\bibnamefont {Wang}}, \bibinfo {author} {\bibfnamefont
  {T.}~\bibnamefont {Brecht}}, \bibinfo {author} {\bibfnamefont {Y.~Y.}\
  \bibnamefont {Gao}}, \bibinfo {author} {\bibfnamefont {S.}~\bibnamefont
  {Shankar}}, \bibinfo {author} {\bibfnamefont {M.}~\bibnamefont {Hatridge}},
  \bibinfo {author} {\bibfnamefont {G.}~\bibnamefont {Catelani}}, \bibinfo
  {author} {\bibfnamefont {M.}~\bibnamefont {Mirrahimi}}, \bibinfo {author}
  {\bibfnamefont {L.}~\bibnamefont {Frunzio}}, \bibinfo {author} {\bibfnamefont
  {R.~J.}\ \bibnamefont {Schoelkopf}}, \bibinfo {author} {\bibfnamefont
  {L.~I.}\ \bibnamefont {Glazman}}, \ and\ \bibinfo {author} {\bibfnamefont
  {M.~H.}\ \bibnamefont {Devoret}},\ }\href@noop {} {\bibfield  {journal}
  {\bibinfo  {journal} {Phys. Rev. Lett.}\ }\textbf {\bibinfo {volume} {113}},\
  \bibinfo {pages} {247001} (\bibinfo {year} {2014})}\BibitemShut {NoStop}%
\bibitem [{\citenamefont {Bell}\ \emph {et~al.}(2012)\citenamefont {Bell},
  \citenamefont {Sadovskyy}, \citenamefont {Ioffe}, \citenamefont {Kitaev},\
  and\ \citenamefont {Gershenson}}]{BellTunableSuperInductanceLoops}%
  \BibitemOpen
  \bibfield  {author} {\bibinfo {author} {\bibfnamefont {M.~T.}\ \bibnamefont
  {Bell}}, \bibinfo {author} {\bibfnamefont {I.~A.}\ \bibnamefont {Sadovskyy}},
  \bibinfo {author} {\bibfnamefont {L.~B.}\ \bibnamefont {Ioffe}}, \bibinfo
  {author} {\bibfnamefont {A.~Y.}\ \bibnamefont {Kitaev}}, \ and\ \bibinfo
  {author} {\bibfnamefont {M.~E.}\ \bibnamefont {Gershenson}},\ }\href@noop {}
  {\bibfield  {journal} {\bibinfo  {journal} {Phys. Rev. Lett.}\ }\textbf
  {\bibinfo {volume} {109}},\ \bibinfo {pages} {137003} (\bibinfo {year}
  {2012})}\BibitemShut {NoStop}%
\bibitem [{\citenamefont {Ioffe}\ \emph {et~al.}(2002)\citenamefont {Ioffe},
  \citenamefont {Feigel'man}, \citenamefont {Ioselevich}, \citenamefont
  {Ivanov}, \citenamefont {Troyer},\ and\ \citenamefont
  {Blatter}}]{IoffeTopologicalQubits2002}%
  \BibitemOpen
  \bibfield  {author} {\bibinfo {author} {\bibfnamefont {L.~B.}\ \bibnamefont
  {Ioffe}}, \bibinfo {author} {\bibfnamefont {M.~V.}\ \bibnamefont
  {Feigel'man}}, \bibinfo {author} {\bibfnamefont {A.}~\bibnamefont
  {Ioselevich}}, \bibinfo {author} {\bibfnamefont {D.}~\bibnamefont {Ivanov}},
  \bibinfo {author} {\bibfnamefont {M.}~\bibnamefont {Troyer}}, \ and\ \bibinfo
  {author} {\bibfnamefont {G.}~\bibnamefont {Blatter}},\ }\href@noop {}
  {\bibfield  {journal} {\bibinfo  {journal} {Nature}\ }\textbf {\bibinfo
  {volume} {415}},\ \bibinfo {pages} {503} (\bibinfo {year}
  {2002})}\BibitemShut {NoStop}%
\bibitem [{\citenamefont {Brooks}\ \emph {et~al.}(2013)\citenamefont {Brooks},
  \citenamefont {Kitaev},\ and\ \citenamefont
  {Preskill}}]{ZeroPiBrooksKiatevPreskill}%
  \BibitemOpen
  \bibfield  {author} {\bibinfo {author} {\bibfnamefont {P.}~\bibnamefont
  {Brooks}}, \bibinfo {author} {\bibfnamefont {A.}~\bibnamefont {Kitaev}}, \
  and\ \bibinfo {author} {\bibfnamefont {J.}~\bibnamefont {Preskill}},\
  }\href@noop {} {\bibfield  {journal} {\bibinfo  {journal} {Phys. Rev. A}\
  }\textbf {\bibinfo {volume} {87}},\ \bibinfo {pages} {052306} (\bibinfo
  {year} {2013})}\BibitemShut {NoStop}%
\bibitem [{\citenamefont {Kitaev}(2006)}]{KitaevZeroPi}%
  \BibitemOpen
  \bibfield  {author} {\bibinfo {author} {\bibfnamefont {A.}~\bibnamefont
  {Kitaev}},\ }\href@noop {} {\bibfield  {journal} {\bibinfo  {journal}
  {arXiv:cond-mat/0609441}\ } (\bibinfo {year} {2006})}\BibitemShut {NoStop}%
\bibitem [{\citenamefont {Earnest}\ \emph {et~al.}(2018)\citenamefont
  {Earnest}, \citenamefont {Chakram}, \citenamefont {Lu}, \citenamefont
  {Irons}, \citenamefont {Naik}, \citenamefont {Leung}, \citenamefont {Ocola},
  \citenamefont {Czaplewski}, \citenamefont {Baker}, \citenamefont {Lawrence},
  \citenamefont {Koch},\ and\ \citenamefont {Schuster}}]{HeavyFluxonium2017}%
  \BibitemOpen
  \bibfield  {author} {\bibinfo {author} {\bibfnamefont {N.}~\bibnamefont
  {Earnest}}, \bibinfo {author} {\bibfnamefont {S.}~\bibnamefont {Chakram}},
  \bibinfo {author} {\bibfnamefont {Y.}~\bibnamefont {Lu}}, \bibinfo {author}
  {\bibfnamefont {N.}~\bibnamefont {Irons}}, \bibinfo {author} {\bibfnamefont
  {R.~K.}\ \bibnamefont {Naik}}, \bibinfo {author} {\bibfnamefont
  {N.}~\bibnamefont {Leung}}, \bibinfo {author} {\bibfnamefont
  {L.}~\bibnamefont {Ocola}}, \bibinfo {author} {\bibfnamefont {D.~A.}\
  \bibnamefont {Czaplewski}}, \bibinfo {author} {\bibfnamefont
  {B.}~\bibnamefont {Baker}}, \bibinfo {author} {\bibfnamefont
  {J.}~\bibnamefont {Lawrence}}, \bibinfo {author} {\bibfnamefont
  {J.}~\bibnamefont {Koch}}, \ and\ \bibinfo {author} {\bibfnamefont {D.~I.}\
  \bibnamefont {Schuster}},\ }\href@noop {} {\bibfield  {journal} {\bibinfo
  {journal} {Phys. Rev. Lett.}\ }\textbf {\bibinfo {volume} {120}},\ \bibinfo
  {pages} {150504} (\bibinfo {year} {2018})}\BibitemShut {NoStop}%
\bibitem [{\citenamefont {Vool}\ \emph {et~al.}(2018)\citenamefont {Vool},
  \citenamefont {Kou}, \citenamefont {Smith}, \citenamefont {Frattini},
  \citenamefont {Serniak}, \citenamefont {Reinhold}, \citenamefont {Pop},
  \citenamefont {Shankar}, \citenamefont {Frunzio}, \citenamefont {Girvin},\
  and\ \citenamefont {Devoret}}]{FluxoniumForbiddenTransitionDriving}%
  \BibitemOpen
  \bibfield  {author} {\bibinfo {author} {\bibfnamefont {U.}~\bibnamefont
  {Vool}}, \bibinfo {author} {\bibfnamefont {A.}~\bibnamefont {Kou}}, \bibinfo
  {author} {\bibfnamefont {W.~C.}\ \bibnamefont {Smith}}, \bibinfo {author}
  {\bibfnamefont {N.~E.}\ \bibnamefont {Frattini}}, \bibinfo {author}
  {\bibfnamefont {K.}~\bibnamefont {Serniak}}, \bibinfo {author} {\bibfnamefont
  {P.}~\bibnamefont {Reinhold}}, \bibinfo {author} {\bibfnamefont {I.~M.}\
  \bibnamefont {Pop}}, \bibinfo {author} {\bibfnamefont {S.}~\bibnamefont
  {Shankar}}, \bibinfo {author} {\bibfnamefont {L.}~\bibnamefont {Frunzio}},
  \bibinfo {author} {\bibfnamefont {S.~M.}\ \bibnamefont {Girvin}}, \ and\
  \bibinfo {author} {\bibfnamefont {M.~H.}\ \bibnamefont {Devoret}},\ }\href
  {\doibase 10.1103/PhysRevApplied.9.054046} {\bibfield  {journal} {\bibinfo
  {journal} {Phys. Rev. Applied}\ }\textbf {\bibinfo {volume} {9}},\ \bibinfo
  {pages} {054046} (\bibinfo {year} {2018})}\BibitemShut {NoStop}%
\bibitem [{\citenamefont {Dempster}\ \emph {et~al.}(2014)\citenamefont
  {Dempster}, \citenamefont {Fu}, \citenamefont {Ferguson}, \citenamefont
  {Schuster},\ and\ \citenamefont {Koch}}]{DempsterZeroPi}%
  \BibitemOpen
  \bibfield  {author} {\bibinfo {author} {\bibfnamefont {J.~M.}\ \bibnamefont
  {Dempster}}, \bibinfo {author} {\bibfnamefont {B.}~\bibnamefont {Fu}},
  \bibinfo {author} {\bibfnamefont {D.~G.}\ \bibnamefont {Ferguson}}, \bibinfo
  {author} {\bibfnamefont {D.~I.}\ \bibnamefont {Schuster}}, \ and\ \bibinfo
  {author} {\bibfnamefont {J.}~\bibnamefont {Koch}},\ }\href@noop {} {\bibfield
   {journal} {\bibinfo  {journal} {Phys. Rev. B}\ }\textbf {\bibinfo {volume}
  {90}},\ \bibinfo {pages} {094518} (\bibinfo {year} {2014})}\BibitemShut
  {NoStop}%
\bibitem [{\citenamefont {Groszkowski}\ \emph {et~al.}(2018)\citenamefont
  {Groszkowski}, \citenamefont {Paolo}, \citenamefont {Grimsmo}, \citenamefont
  {Blais}, \citenamefont {Schuster}, \citenamefont {Houck},\ and\ \citenamefont
  {Koch}}]{PeterZeroPi2017}%
  \BibitemOpen
  \bibfield  {author} {\bibinfo {author} {\bibfnamefont {P.}~\bibnamefont
  {Groszkowski}}, \bibinfo {author} {\bibfnamefont {A.~D.}\ \bibnamefont
  {Paolo}}, \bibinfo {author} {\bibfnamefont {A.~L.}\ \bibnamefont {Grimsmo}},
  \bibinfo {author} {\bibfnamefont {A.}~\bibnamefont {Blais}}, \bibinfo
  {author} {\bibfnamefont {D.~I.}\ \bibnamefont {Schuster}}, \bibinfo {author}
  {\bibfnamefont {A.~A.}\ \bibnamefont {Houck}}, \ and\ \bibinfo {author}
  {\bibfnamefont {J.}~\bibnamefont {Koch}},\ }\href
  {http://stacks.iop.org/1367-2630/20/i=4/a=043053} {\bibfield  {journal}
  {\bibinfo  {journal} {New Journal of Physics}\ }\textbf {\bibinfo {volume}
  {20}},\ \bibinfo {pages} {043053} (\bibinfo {year} {2018})}\BibitemShut
  {NoStop}%
\bibitem [{\citenamefont {Kerman}\ \emph {et~al.}(2007)\citenamefont {Kerman},
  \citenamefont {Dauler}, \citenamefont {Yang}, \citenamefont {Rosfjord},
  \citenamefont {Anant}, \citenamefont {Berggren}, \citenamefont
  {Gol’tsman},\ and\ \citenamefont {Voronov}}]{KermanNbNdetectors2007}%
  \BibitemOpen
  \bibfield  {author} {\bibinfo {author} {\bibfnamefont {A.~J.}\ \bibnamefont
  {Kerman}}, \bibinfo {author} {\bibfnamefont {E.~A.}\ \bibnamefont {Dauler}},
  \bibinfo {author} {\bibfnamefont {J.~K.~W.}\ \bibnamefont {Yang}}, \bibinfo
  {author} {\bibfnamefont {K.~M.}\ \bibnamefont {Rosfjord}}, \bibinfo {author}
  {\bibfnamefont {V.}~\bibnamefont {Anant}}, \bibinfo {author} {\bibfnamefont
  {K.~K.}\ \bibnamefont {Berggren}}, \bibinfo {author} {\bibfnamefont {G.~N.}\
  \bibnamefont {Gol’tsman}}, \ and\ \bibinfo {author} {\bibfnamefont {B.~M.}\
  \bibnamefont {Voronov}},\ }\href@noop {} {\bibfield  {journal} {\bibinfo
  {journal} {Applied Physics Letters}\ }\textbf {\bibinfo {volume} {90}},\
  \bibinfo {pages} {101110} (\bibinfo {year} {2007})}\BibitemShut {NoStop}%
\bibitem [{\citenamefont {Annunziata}\ \emph {et~al.}(2010)\citenamefont
  {Annunziata}, \citenamefont {Santavicca}, \citenamefont {Frunzio},
  \citenamefont {Catelani}, \citenamefont {Rooks}, \citenamefont {Frydman},\
  and\ \citenamefont {Prober}}]{tunableSuperInductance2010}%
  \BibitemOpen
  \bibfield  {author} {\bibinfo {author} {\bibfnamefont {A.~J.}\ \bibnamefont
  {Annunziata}}, \bibinfo {author} {\bibfnamefont {D.~F.}\ \bibnamefont
  {Santavicca}}, \bibinfo {author} {\bibfnamefont {L.}~\bibnamefont {Frunzio}},
  \bibinfo {author} {\bibfnamefont {G.}~\bibnamefont {Catelani}}, \bibinfo
  {author} {\bibfnamefont {M.~J.}\ \bibnamefont {Rooks}}, \bibinfo {author}
  {\bibfnamefont {A.}~\bibnamefont {Frydman}}, \ and\ \bibinfo {author}
  {\bibfnamefont {D.~E.}\ \bibnamefont {Prober}},\ }\href@noop {} {\bibfield
  {journal} {\bibinfo  {journal} {Nanotechnology}\ }\textbf {\bibinfo {volume}
  {21}},\ \bibinfo {pages} {445202} (\bibinfo {year} {2010})}\BibitemShut
  {NoStop}%
\bibitem [{\citenamefont {Santavicca}\ \emph {et~al.}(2016)\citenamefont
  {Santavicca}, \citenamefont {Adams}, \citenamefont {Grant}, \citenamefont
  {McCaughan},\ and\ \citenamefont {Berggren}}]{BerggrenNbNresonators2016}%
  \BibitemOpen
  \bibfield  {author} {\bibinfo {author} {\bibfnamefont {D.~F.}\ \bibnamefont
  {Santavicca}}, \bibinfo {author} {\bibfnamefont {J.~K.}\ \bibnamefont
  {Adams}}, \bibinfo {author} {\bibfnamefont {L.~E.}\ \bibnamefont {Grant}},
  \bibinfo {author} {\bibfnamefont {A.~N.}\ \bibnamefont {McCaughan}}, \ and\
  \bibinfo {author} {\bibfnamefont {K.~K.}\ \bibnamefont {Berggren}},\
  }\href@noop {} {\bibfield  {journal} {\bibinfo  {journal} {Journal of Applied
  Physics}\ }\textbf {\bibinfo {volume} {119}},\ \bibinfo {pages} {234302}
  (\bibinfo {year} {2016})}\BibitemShut {NoStop}%
\bibitem [{\citenamefont {Ho~Eom}\ \emph {et~al.}(2012)\citenamefont {Ho~Eom},
  \citenamefont {Day}, \citenamefont {LeDuc},\ and\ \citenamefont
  {Zmuidzinas}}]{KineticInductanceParaamp2012}%
  \BibitemOpen
  \bibfield  {author} {\bibinfo {author} {\bibfnamefont {B.}~\bibnamefont
  {Ho~Eom}}, \bibinfo {author} {\bibfnamefont {P.~K.}\ \bibnamefont {Day}},
  \bibinfo {author} {\bibfnamefont {H.~G.}\ \bibnamefont {LeDuc}}, \ and\
  \bibinfo {author} {\bibfnamefont {J.}~\bibnamefont {Zmuidzinas}},\
  }\href@noop {} {\bibfield  {journal} {\bibinfo  {journal} {Nature Physics}\
  }\textbf {\bibinfo {volume} {8}},\ \bibinfo {pages} {623} (\bibinfo {year}
  {2012})}\BibitemShut {NoStop}%
\bibitem [{\citenamefont {Vissers}\ \emph {et~al.}(2016)\citenamefont
  {Vissers}, \citenamefont {Erickson}, \citenamefont {Ku}, \citenamefont
  {Vale}, \citenamefont {Wu}, \citenamefont {Hilton},\ and\ \citenamefont
  {Pappas}}]{NbTiNparaAmp2016}%
  \BibitemOpen
  \bibfield  {author} {\bibinfo {author} {\bibfnamefont {M.~R.}\ \bibnamefont
  {Vissers}}, \bibinfo {author} {\bibfnamefont {R.~P.}\ \bibnamefont
  {Erickson}}, \bibinfo {author} {\bibfnamefont {H.-S.}\ \bibnamefont {Ku}},
  \bibinfo {author} {\bibfnamefont {L.}~\bibnamefont {Vale}}, \bibinfo {author}
  {\bibfnamefont {X.}~\bibnamefont {Wu}}, \bibinfo {author} {\bibfnamefont
  {G.~C.}\ \bibnamefont {Hilton}}, \ and\ \bibinfo {author} {\bibfnamefont
  {D.~P.}\ \bibnamefont {Pappas}},\ }\href@noop {} {\bibfield  {journal}
  {\bibinfo  {journal} {Applied Physics Letters}\ }\textbf {\bibinfo {volume}
  {108}},\ \bibinfo {pages} {012601} (\bibinfo {year} {2016})}\BibitemShut
  {NoStop}%
\bibitem [{\citenamefont {Chaudhuri}\ \emph {et~al.}(2017)\citenamefont
  {Chaudhuri}, \citenamefont {Li}, \citenamefont {Irwin}, \citenamefont
  {Bockstiegel}, \citenamefont {Hubmayr}, \citenamefont {Ullom}, \citenamefont
  {Vissers},\ and\ \citenamefont {Gao}}]{KITWPA2017}%
  \BibitemOpen
  \bibfield  {author} {\bibinfo {author} {\bibfnamefont {S.}~\bibnamefont
  {Chaudhuri}}, \bibinfo {author} {\bibfnamefont {D.}~\bibnamefont {Li}},
  \bibinfo {author} {\bibfnamefont {K.~D.}\ \bibnamefont {Irwin}}, \bibinfo
  {author} {\bibfnamefont {C.}~\bibnamefont {Bockstiegel}}, \bibinfo {author}
  {\bibfnamefont {J.}~\bibnamefont {Hubmayr}}, \bibinfo {author} {\bibfnamefont
  {J.~N.}\ \bibnamefont {Ullom}}, \bibinfo {author} {\bibfnamefont {M.~R.}\
  \bibnamefont {Vissers}}, \ and\ \bibinfo {author} {\bibfnamefont
  {J.}~\bibnamefont {Gao}},\ }\href@noop {} {\bibfield  {journal} {\bibinfo
  {journal} {Applied Physics Letters}\ }\textbf {\bibinfo {volume} {110}},\
  \bibinfo {pages} {152601} (\bibinfo {year} {2017})}\BibitemShut {NoStop}%
\bibitem [{\citenamefont {Kerman}(2010)}]{KermanNanoWireQubitPRL2010}%
  \BibitemOpen
  \bibfield  {author} {\bibinfo {author} {\bibfnamefont {A.~J.}\ \bibnamefont
  {Kerman}},\ }\href@noop {} {\bibfield  {journal} {\bibinfo  {journal} {Phys.
  Rev. Lett.}\ }\textbf {\bibinfo {volume} {104}},\ \bibinfo {pages} {027002}
  (\bibinfo {year} {2010})}\BibitemShut {NoStop}%
\bibitem [{\citenamefont {Peltonen}\ \emph {et~al.}(2018)\citenamefont
  {Peltonen}, \citenamefont {Coumou}, \citenamefont {Peng}, \citenamefont
  {Klapwijk}, \citenamefont {Tsai},\ and\ \citenamefont
  {Astafiev}}]{rFSquidKineticIndcutance2017}%
  \BibitemOpen
  \bibfield  {author} {\bibinfo {author} {\bibfnamefont {J.~T.}\ \bibnamefont
  {Peltonen}}, \bibinfo {author} {\bibfnamefont {P.~C. J.~J.}\ \bibnamefont
  {Coumou}}, \bibinfo {author} {\bibfnamefont {Z.~H.}\ \bibnamefont {Peng}},
  \bibinfo {author} {\bibfnamefont {T.~M.}\ \bibnamefont {Klapwijk}}, \bibinfo
  {author} {\bibfnamefont {J.~S.}\ \bibnamefont {Tsai}}, \ and\ \bibinfo
  {author} {\bibfnamefont {O.~V.}\ \bibnamefont {Astafiev}},\ }\href
  {https://doi.org/10.1038/s41598-018-27154-1} {\bibfield  {journal} {\bibinfo
  {journal} {Scientific Reports}\ }\textbf {\bibinfo {volume} {8}},\ \bibinfo
  {pages} {10033} (\bibinfo {year} {2018})}\BibitemShut {NoStop}%
\bibitem [{\citenamefont {Tinkham}(1996)}]{Tinkham}%
  \BibitemOpen
  \bibfield  {author} {\bibinfo {author} {\bibfnamefont {M.}~\bibnamefont
  {Tinkham}},\ }\href@noop {} {\emph {\bibinfo {title} {Introduction to
  Superconductivity, 2nd Ed.}}}\ (\bibinfo  {publisher} {McGraw-Hill Book
  Co.},\ \bibinfo {year} {1996})\BibitemShut {NoStop}%
\bibitem [{\citenamefont {Viola}\ and\ \citenamefont
  {Catelani}(2015)}]{FluxoniumCollectiveModes2015}%
  \BibitemOpen
  \bibfield  {author} {\bibinfo {author} {\bibfnamefont {G.}~\bibnamefont
  {Viola}}\ and\ \bibinfo {author} {\bibfnamefont {G.}~\bibnamefont
  {Catelani}},\ }\href@noop {} {\bibfield  {journal} {\bibinfo  {journal}
  {Phys. Rev. B}\ }\textbf {\bibinfo {volume} {92}},\ \bibinfo {pages} {224511}
  (\bibinfo {year} {2015})}\BibitemShut {NoStop}%
\bibitem [{\citenamefont {Ferguson}\ \emph {et~al.}(2013)\citenamefont
  {Ferguson}, \citenamefont {Houck},\ and\ \citenamefont
  {Koch}}]{JensCollectiveExcitations2013}%
  \BibitemOpen
  \bibfield  {author} {\bibinfo {author} {\bibfnamefont {D.~G.}\ \bibnamefont
  {Ferguson}}, \bibinfo {author} {\bibfnamefont {A.~A.}\ \bibnamefont {Houck}},
  \ and\ \bibinfo {author} {\bibfnamefont {J.}~\bibnamefont {Koch}},\
  }\href@noop {} {\bibfield  {journal} {\bibinfo  {journal} {Phys. Rev. X}\
  }\textbf {\bibinfo {volume} {3}},\ \bibinfo {pages} {011003} (\bibinfo {year}
  {2013})}\BibitemShut {NoStop}%
\bibitem [{\citenamefont {Nigg}\ \emph {et~al.}(2012)\citenamefont {Nigg},
  \citenamefont {Paik}, \citenamefont {Vlastakis}, \citenamefont {Kirchmair},
  \citenamefont {Shankar}, \citenamefont {Frunzio}, \citenamefont {Devoret},
  \citenamefont {Schoelkopf},\ and\ \citenamefont {Girvin}}]{BBQPRL2012}%
  \BibitemOpen
  \bibfield  {author} {\bibinfo {author} {\bibfnamefont {S.~E.}\ \bibnamefont
  {Nigg}}, \bibinfo {author} {\bibfnamefont {H.}~\bibnamefont {Paik}}, \bibinfo
  {author} {\bibfnamefont {B.}~\bibnamefont {Vlastakis}}, \bibinfo {author}
  {\bibfnamefont {G.}~\bibnamefont {Kirchmair}}, \bibinfo {author}
  {\bibfnamefont {S.}~\bibnamefont {Shankar}}, \bibinfo {author} {\bibfnamefont
  {L.}~\bibnamefont {Frunzio}}, \bibinfo {author} {\bibfnamefont {M.~H.}\
  \bibnamefont {Devoret}}, \bibinfo {author} {\bibfnamefont {R.~J.}\
  \bibnamefont {Schoelkopf}}, \ and\ \bibinfo {author} {\bibfnamefont {S.~M.}\
  \bibnamefont {Girvin}},\ }\href {\doibase 10.1103/PhysRevLett.108.240502}
  {\bibfield  {journal} {\bibinfo  {journal} {Phys. Rev. Lett.}\ }\textbf
  {\bibinfo {volume} {108}},\ \bibinfo {pages} {240502} (\bibinfo {year}
  {2012})}\BibitemShut {NoStop}%
\bibitem [{\citenamefont {Solgun}\ \emph {et~al.}(2014)\citenamefont {Solgun},
  \citenamefont {Abraham},\ and\ \citenamefont
  {DiVincenzo}}]{solgun2014blackbox}%
  \BibitemOpen
  \bibfield  {author} {\bibinfo {author} {\bibfnamefont {F.}~\bibnamefont
  {Solgun}}, \bibinfo {author} {\bibfnamefont {D.~W.}\ \bibnamefont {Abraham}},
  \ and\ \bibinfo {author} {\bibfnamefont {D.~P.}\ \bibnamefont {DiVincenzo}},\
  }\href@noop {} {\bibfield  {journal} {\bibinfo  {journal} {Phys. Rev. B}\
  }\textbf {\bibinfo {volume} {90}},\ \bibinfo {pages} {134504} (\bibinfo
  {year} {2014})}\BibitemShut {NoStop}%
\bibitem [{\citenamefont {Koch}\ \emph {et~al.}(2009)\citenamefont {Koch},
  \citenamefont {Manucharyan}, \citenamefont {Devoret},\ and\ \citenamefont
  {Glazman}}]{ChargingInFluxonium2009}%
  \BibitemOpen
  \bibfield  {author} {\bibinfo {author} {\bibfnamefont {J.}~\bibnamefont
  {Koch}}, \bibinfo {author} {\bibfnamefont {V.}~\bibnamefont {Manucharyan}},
  \bibinfo {author} {\bibfnamefont {M.~H.}\ \bibnamefont {Devoret}}, \ and\
  \bibinfo {author} {\bibfnamefont {L.~I.}\ \bibnamefont {Glazman}},\
  }\href@noop {} {\bibfield  {journal} {\bibinfo  {journal} {Phys. Rev. Lett.}\
  }\textbf {\bibinfo {volume} {103}},\ \bibinfo {pages} {217004} (\bibinfo
  {year} {2009})}\BibitemShut {NoStop}%
\bibitem [{\citenamefont {Manucharyan}\ \emph {et~al.}(2012)\citenamefont
  {Manucharyan}, \citenamefont {Masluk}, \citenamefont {Kamal}, \citenamefont
  {Koch}, \citenamefont {Glazman},\ and\ \citenamefont
  {Devoret}}]{QCPSFluxoniumValdimir2012}%
  \BibitemOpen
  \bibfield  {author} {\bibinfo {author} {\bibfnamefont {V.~E.}\ \bibnamefont
  {Manucharyan}}, \bibinfo {author} {\bibfnamefont {N.~A.}\ \bibnamefont
  {Masluk}}, \bibinfo {author} {\bibfnamefont {A.}~\bibnamefont {Kamal}},
  \bibinfo {author} {\bibfnamefont {J.}~\bibnamefont {Koch}}, \bibinfo {author}
  {\bibfnamefont {L.~I.}\ \bibnamefont {Glazman}}, \ and\ \bibinfo {author}
  {\bibfnamefont {M.~H.}\ \bibnamefont {Devoret}},\ }\href@noop {} {\bibfield
  {journal} {\bibinfo  {journal} {Phys. Rev. B}\ }\textbf {\bibinfo {volume}
  {85}},\ \bibinfo {pages} {024521} (\bibinfo {year} {2012})}\BibitemShut
  {NoStop}%
\bibitem [{\citenamefont {Zhu}\ \emph {et~al.}(2013)\citenamefont {Zhu},
  \citenamefont {Ferguson}, \citenamefont {Manucharyan},\ and\ \citenamefont
  {Koch}}]{JensDisperssiveFluxonium2013}%
  \BibitemOpen
  \bibfield  {author} {\bibinfo {author} {\bibfnamefont {G.}~\bibnamefont
  {Zhu}}, \bibinfo {author} {\bibfnamefont {D.~G.}\ \bibnamefont {Ferguson}},
  \bibinfo {author} {\bibfnamefont {V.~E.}\ \bibnamefont {Manucharyan}}, \ and\
  \bibinfo {author} {\bibfnamefont {J.}~\bibnamefont {Koch}},\ }\href@noop {}
  {\bibfield  {journal} {\bibinfo  {journal} {Phys. Rev. B}\ }\textbf {\bibinfo
  {volume} {87}},\ \bibinfo {pages} {024510} (\bibinfo {year}
  {2013})}\BibitemShut {NoStop}%
\bibitem [{\citenamefont {Lin}\ \emph {et~al.}(2018)\citenamefont {Lin},
  \citenamefont {Nguyen}, \citenamefont {Grabon}, \citenamefont {San~Miguel},
  \citenamefont {Pankratova},\ and\ \citenamefont
  {Manucharyan}}]{FluxoniumSelectrionRuleEngineering2018}%
  \BibitemOpen
  \bibfield  {author} {\bibinfo {author} {\bibfnamefont {Y.-H.}\ \bibnamefont
  {Lin}}, \bibinfo {author} {\bibfnamefont {L.~B.}\ \bibnamefont {Nguyen}},
  \bibinfo {author} {\bibfnamefont {N.}~\bibnamefont {Grabon}}, \bibinfo
  {author} {\bibfnamefont {J.}~\bibnamefont {San~Miguel}}, \bibinfo {author}
  {\bibfnamefont {N.}~\bibnamefont {Pankratova}}, \ and\ \bibinfo {author}
  {\bibfnamefont {V.~E.}\ \bibnamefont {Manucharyan}},\ }\href@noop {}
  {\bibfield  {journal} {\bibinfo  {journal} {Phys. Rev. Lett.}\ }\textbf
  {\bibinfo {volume} {120}},\ \bibinfo {pages} {150503} (\bibinfo {year}
  {2018})}\BibitemShut {NoStop}%
\bibitem [{\citenamefont {material}()}]{Supplement}%
  \BibitemOpen
  \bibfield  {author} {\bibinfo {author} {\bibfnamefont {}\ \bibnamefont
  {See supplemental material}}}\href@noop {} {\ }\BibitemShut {NoStop}%
\bibitem [{\citenamefont {Wallraff}\ \emph {et~al.}(2007)\citenamefont
  {Wallraff}, \citenamefont {Schuster}, \citenamefont {Blais}, \citenamefont
  {Gambetta}, \citenamefont {Schreier}, \citenamefont {Frunzio}, \citenamefont
  {Devoret}, \citenamefont {Girvin},\ and\ \citenamefont
  {Schoelkopf}}]{WallraffSideBandTransitions2007}%
  \BibitemOpen
  \bibfield  {author} {\bibinfo {author} {\bibfnamefont {A.}~\bibnamefont
  {Wallraff}}, \bibinfo {author} {\bibfnamefont {D.~I.}\ \bibnamefont
  {Schuster}}, \bibinfo {author} {\bibfnamefont {A.}~\bibnamefont {Blais}},
  \bibinfo {author} {\bibfnamefont {J.~M.}\ \bibnamefont {Gambetta}}, \bibinfo
  {author} {\bibfnamefont {J.}~\bibnamefont {Schreier}}, \bibinfo {author}
  {\bibfnamefont {L.}~\bibnamefont {Frunzio}}, \bibinfo {author} {\bibfnamefont
  {M.~H.}\ \bibnamefont {Devoret}}, \bibinfo {author} {\bibfnamefont {S.~M.}\
  \bibnamefont {Girvin}}, \ and\ \bibinfo {author} {\bibfnamefont {R.~J.}\
  \bibnamefont {Schoelkopf}},\ }\href@noop {} {\bibfield  {journal} {\bibinfo
  {journal} {Phys. Rev. Lett.}\ }\textbf {\bibinfo {volume} {99}},\ \bibinfo
  {pages} {050501} (\bibinfo {year} {2007})}\BibitemShut {NoStop}%
\bibitem [{\citenamefont {Bourassa}\ \emph {et~al.}(2012)\citenamefont
  {Bourassa}, \citenamefont {Beaudoin}, \citenamefont {Gambetta},\ and\
  \citenamefont {Blais}}]{JJembededInTXLine2012}%
  \BibitemOpen
  \bibfield  {author} {\bibinfo {author} {\bibfnamefont {J.}~\bibnamefont
  {Bourassa}}, \bibinfo {author} {\bibfnamefont {F.}~\bibnamefont {Beaudoin}},
  \bibinfo {author} {\bibfnamefont {J.~M.}\ \bibnamefont {Gambetta}}, \ and\
  \bibinfo {author} {\bibfnamefont {A.}~\bibnamefont {Blais}},\ }\href@noop {}
  {\bibfield  {journal} {\bibinfo  {journal} {Phys. Rev. A}\ }\textbf {\bibinfo
  {volume} {86}},\ \bibinfo {pages} {013814} (\bibinfo {year}
  {2012})}\BibitemShut {NoStop}%
\bibitem [{\citenamefont {Mortensen}\ \emph {et~al.}(2016)\citenamefont
  {Mortensen}, \citenamefont {M\o{}lmer},\ and\ \citenamefont
  {Andersen}}]{NormalModesTXResonator2016}%
  \BibitemOpen
  \bibfield  {author} {\bibinfo {author} {\bibfnamefont {H.~L.}\ \bibnamefont
  {Mortensen}}, \bibinfo {author} {\bibfnamefont {K.}~\bibnamefont
  {M\o{}lmer}}, \ and\ \bibinfo {author} {\bibfnamefont {C.~K.}\ \bibnamefont
  {Andersen}},\ }\href@noop {} {\bibfield  {journal} {\bibinfo  {journal}
  {Phys. Rev. A}\ }\textbf {\bibinfo {volume} {94}},\ \bibinfo {pages} {053817}
  (\bibinfo {year} {2016})}\BibitemShut {NoStop}%
\bibitem [{\citenamefont {Baur}\ \emph {et~al.}(2009)\citenamefont {Baur},
  \citenamefont {Filipp}, \citenamefont {Bianchetti}, \citenamefont {Fink},
  \citenamefont {G\"oppl}, \citenamefont {Steffen}, \citenamefont {Leek},
  \citenamefont {Blais},\ and\ \citenamefont
  {Wallraff}}]{WallraffATsplitting2009}%
  \BibitemOpen
  \bibfield  {author} {\bibinfo {author} {\bibfnamefont {M.}~\bibnamefont
  {Baur}}, \bibinfo {author} {\bibfnamefont {S.}~\bibnamefont {Filipp}},
  \bibinfo {author} {\bibfnamefont {R.}~\bibnamefont {Bianchetti}}, \bibinfo
  {author} {\bibfnamefont {J.~M.}\ \bibnamefont {Fink}}, \bibinfo {author}
  {\bibfnamefont {M.}~\bibnamefont {G\"oppl}}, \bibinfo {author} {\bibfnamefont
  {L.}~\bibnamefont {Steffen}}, \bibinfo {author} {\bibfnamefont {P.~J.}\
  \bibnamefont {Leek}}, \bibinfo {author} {\bibfnamefont {A.}~\bibnamefont
  {Blais}}, \ and\ \bibinfo {author} {\bibfnamefont {A.}~\bibnamefont
  {Wallraff}},\ }\href@noop {} {\bibfield  {journal} {\bibinfo  {journal}
  {Phys. Rev. Lett.}\ }\textbf {\bibinfo {volume} {102}},\ \bibinfo {pages}
  {243602} (\bibinfo {year} {2009})}\BibitemShut {NoStop}%
\bibitem [{\citenamefont {Sillanp\"a\"a}\ \emph {et~al.}(2009)\citenamefont
  {Sillanp\"a\"a}, \citenamefont {Li}, \citenamefont {Cicak}, \citenamefont
  {Altomare}, \citenamefont {Park}, \citenamefont {Simmonds}, \citenamefont
  {Paraoanu},\ and\ \citenamefont {Hakonen}}]{HakonenAT3levels2009}%
  \BibitemOpen
  \bibfield  {author} {\bibinfo {author} {\bibfnamefont {M.~A.}\ \bibnamefont
  {Sillanp\"a\"a}}, \bibinfo {author} {\bibfnamefont {J.}~\bibnamefont {Li}},
  \bibinfo {author} {\bibfnamefont {K.}~\bibnamefont {Cicak}}, \bibinfo
  {author} {\bibfnamefont {F.}~\bibnamefont {Altomare}}, \bibinfo {author}
  {\bibfnamefont {J.~I.}\ \bibnamefont {Park}}, \bibinfo {author}
  {\bibfnamefont {R.~W.}\ \bibnamefont {Simmonds}}, \bibinfo {author}
  {\bibfnamefont {G.~S.}\ \bibnamefont {Paraoanu}}, \ and\ \bibinfo {author}
  {\bibfnamefont {P.~J.}\ \bibnamefont {Hakonen}},\ }\href@noop {} {\bibfield
  {journal} {\bibinfo  {journal} {Phys. Rev. Lett.}\ }\textbf {\bibinfo
  {volume} {103}},\ \bibinfo {pages} {193601} (\bibinfo {year}
  {2009})}\BibitemShut {NoStop}%
\bibitem [{\citenamefont {Novikov}\ \emph {et~al.}(2013)\citenamefont
  {Novikov}, \citenamefont {Robinson}, \citenamefont {Keane}, \citenamefont
  {Suri}, \citenamefont {Wellstood},\ and\ \citenamefont
  {Palmer}}]{AT3DtransmonPalmer}%
  \BibitemOpen
  \bibfield  {author} {\bibinfo {author} {\bibfnamefont {S.}~\bibnamefont
  {Novikov}}, \bibinfo {author} {\bibfnamefont {J.~E.}\ \bibnamefont
  {Robinson}}, \bibinfo {author} {\bibfnamefont {Z.~K.}\ \bibnamefont {Keane}},
  \bibinfo {author} {\bibfnamefont {B.}~\bibnamefont {Suri}}, \bibinfo {author}
  {\bibfnamefont {F.~C.}\ \bibnamefont {Wellstood}}, \ and\ \bibinfo {author}
  {\bibfnamefont {B.~S.}\ \bibnamefont {Palmer}},\ }\href@noop {} {\bibfield
  {journal} {\bibinfo  {journal} {Phys. Rev. B}\ }\textbf {\bibinfo {volume}
  {88}},\ \bibinfo {pages} {060503} (\bibinfo {year} {2013})}\BibitemShut
  {NoStop}%
\bibitem [{\citenamefont {Suri}\ \emph {et~al.}(2013)\citenamefont {Suri},
  \citenamefont {Keane}, \citenamefont {Ruskov}, \citenamefont {Bishop},
  \citenamefont {Tahan}, \citenamefont {Novikov}, \citenamefont {Robinson},
  \citenamefont {Wellstood},\ and\ \citenamefont {Palmer}}]{PalmerATNJP2013}%
  \BibitemOpen
  \bibfield  {author} {\bibinfo {author} {\bibfnamefont {B.}~\bibnamefont
  {Suri}}, \bibinfo {author} {\bibfnamefont {Z.~K.}\ \bibnamefont {Keane}},
  \bibinfo {author} {\bibfnamefont {R.}~\bibnamefont {Ruskov}}, \bibinfo
  {author} {\bibfnamefont {L.~S.}\ \bibnamefont {Bishop}}, \bibinfo {author}
  {\bibfnamefont {C.}~\bibnamefont {Tahan}}, \bibinfo {author} {\bibfnamefont
  {S.}~\bibnamefont {Novikov}}, \bibinfo {author} {\bibfnamefont {J.~E.}\
  \bibnamefont {Robinson}}, \bibinfo {author} {\bibfnamefont {F.~C.}\
  \bibnamefont {Wellstood}}, \ and\ \bibinfo {author} {\bibfnamefont {B.~S.}\
  \bibnamefont {Palmer}},\ }\href@noop {} {\bibfield  {journal} {\bibinfo
  {journal} {New Journal of Physics}\ }\textbf {\bibinfo {volume} {15}},\
  \bibinfo {pages} {125007} (\bibinfo {year} {2013})}\BibitemShut {NoStop}%
\bibitem [{\citenamefont {Peterer}\ \emph {et~al.}(2015)\citenamefont
  {Peterer}, \citenamefont {Bader}, \citenamefont {Jin}, \citenamefont {Yan},
  \citenamefont {Kamal}, \citenamefont {Gudmundsen}, \citenamefont {Leek},
  \citenamefont {Orlando}, \citenamefont {Oliver},\ and\ \citenamefont
  {Gustavsson}}]{MultilevelPiPulse2015}%
  \BibitemOpen
  \bibfield  {author} {\bibinfo {author} {\bibfnamefont {M.~J.}\ \bibnamefont
  {Peterer}}, \bibinfo {author} {\bibfnamefont {S.~J.}\ \bibnamefont {Bader}},
  \bibinfo {author} {\bibfnamefont {X.}~\bibnamefont {Jin}}, \bibinfo {author}
  {\bibfnamefont {F.}~\bibnamefont {Yan}}, \bibinfo {author} {\bibfnamefont
  {A.}~\bibnamefont {Kamal}}, \bibinfo {author} {\bibfnamefont {T.~J.}\
  \bibnamefont {Gudmundsen}}, \bibinfo {author} {\bibfnamefont {P.~J.}\
  \bibnamefont {Leek}}, \bibinfo {author} {\bibfnamefont {T.~P.}\ \bibnamefont
  {Orlando}}, \bibinfo {author} {\bibfnamefont {W.~D.}\ \bibnamefont {Oliver}},
  \ and\ \bibinfo {author} {\bibfnamefont {S.}~\bibnamefont {Gustavsson}},\
  }\href@noop {} {\bibfield  {journal} {\bibinfo  {journal} {Phys. Rev. Lett.}\
  }\textbf {\bibinfo {volume} {114}},\ \bibinfo {pages} {010501} (\bibinfo
  {year} {2015})}\BibitemShut {NoStop}%
\bibitem [{\citenamefont {Schoelkopf}\ \emph {et~al.}(2002)\citenamefont
  {Schoelkopf}, \citenamefont {Clerk}, \citenamefont {Girvin}, \citenamefont
  {Lehnert},\ and\ \citenamefont {Devoret}}]{SchoelkopfQubitsAsNoiseSpec2002}%
  \BibitemOpen
  \bibfield  {author} {\bibinfo {author} {\bibfnamefont {R.~J.}\ \bibnamefont
  {Schoelkopf}}, \bibinfo {author} {\bibfnamefont {A.~A.}\ \bibnamefont
  {Clerk}}, \bibinfo {author} {\bibfnamefont {S.~M.}\ \bibnamefont {Girvin}},
  \bibinfo {author} {\bibfnamefont {K.~W.}\ \bibnamefont {Lehnert}}, \ and\
  \bibinfo {author} {\bibfnamefont {M.~H.}\ \bibnamefont {Devoret}},\
  }\href@noop {} {\bibfield  {journal} {\bibinfo  {journal}
  {arXiv:cond-mat/0210247}\ } (\bibinfo {year} {2002})}\BibitemShut {NoStop}%
\bibitem [{\citenamefont {Houck}\ \emph {et~al.}(2008)\citenamefont {Houck},
  \citenamefont {Schreier}, \citenamefont {Johnson}, \citenamefont {Chow},
  \citenamefont {Koch}, \citenamefont {Gambetta}, \citenamefont {Schuster},
  \citenamefont {Frunzio}, \citenamefont {Devoret}, \citenamefont {Girvin},\
  and\ \citenamefont {Schoelkopf}}]{AndrewSpontaneousEmission}%
  \BibitemOpen
  \bibfield  {author} {\bibinfo {author} {\bibfnamefont {A.~A.}\ \bibnamefont
  {Houck}}, \bibinfo {author} {\bibfnamefont {J.~A.}\ \bibnamefont {Schreier}},
  \bibinfo {author} {\bibfnamefont {B.~R.}\ \bibnamefont {Johnson}}, \bibinfo
  {author} {\bibfnamefont {J.~M.}\ \bibnamefont {Chow}}, \bibinfo {author}
  {\bibfnamefont {J.}~\bibnamefont {Koch}}, \bibinfo {author} {\bibfnamefont
  {J.~M.}\ \bibnamefont {Gambetta}}, \bibinfo {author} {\bibfnamefont {D.~I.}\
  \bibnamefont {Schuster}}, \bibinfo {author} {\bibfnamefont {L.}~\bibnamefont
  {Frunzio}}, \bibinfo {author} {\bibfnamefont {M.~H.}\ \bibnamefont
  {Devoret}}, \bibinfo {author} {\bibfnamefont {S.~M.}\ \bibnamefont {Girvin}},
  \ and\ \bibinfo {author} {\bibfnamefont {R.~J.}\ \bibnamefont {Schoelkopf}},\
  }\href@noop {} {\bibfield  {journal} {\bibinfo  {journal} {Phys. Rev. Lett.}\
  }\textbf {\bibinfo {volume} {101}},\ \bibinfo {pages} {080502} (\bibinfo
  {year} {2008})}\BibitemShut {NoStop}%
\bibitem [{\citenamefont {Paik}\ \emph {et~al.}(2011)\citenamefont {Paik},
  \citenamefont {Schuster}, \citenamefont {Bishop}, \citenamefont {Kirchmair},
  \citenamefont {Catelani}, \citenamefont {Sears}, \citenamefont {Johnson},
  \citenamefont {Reagor}, \citenamefont {Frunzio}, \citenamefont {Glazman},
  \citenamefont {Girvin}, \citenamefont {Devoret},\ and\ \citenamefont
  {Schoelkopf}}]{3DtmonYale2011}%
  \BibitemOpen
  \bibfield  {author} {\bibinfo {author} {\bibfnamefont {H.}~\bibnamefont
  {Paik}}, \bibinfo {author} {\bibfnamefont {D.~I.}\ \bibnamefont {Schuster}},
  \bibinfo {author} {\bibfnamefont {L.~S.}\ \bibnamefont {Bishop}}, \bibinfo
  {author} {\bibfnamefont {G.}~\bibnamefont {Kirchmair}}, \bibinfo {author}
  {\bibfnamefont {G.}~\bibnamefont {Catelani}}, \bibinfo {author}
  {\bibfnamefont {A.~P.}\ \bibnamefont {Sears}}, \bibinfo {author}
  {\bibfnamefont {B.~R.}\ \bibnamefont {Johnson}}, \bibinfo {author}
  {\bibfnamefont {M.~J.}\ \bibnamefont {Reagor}}, \bibinfo {author}
  {\bibfnamefont {L.}~\bibnamefont {Frunzio}}, \bibinfo {author} {\bibfnamefont
  {L.~I.}\ \bibnamefont {Glazman}}, \bibinfo {author} {\bibfnamefont {S.~M.}\
  \bibnamefont {Girvin}}, \bibinfo {author} {\bibfnamefont {M.~H.}\
  \bibnamefont {Devoret}}, \ and\ \bibinfo {author} {\bibfnamefont {R.~J.}\
  \bibnamefont {Schoelkopf}},\ }\href {\doibase 10.1103/PhysRevLett.107.240501}
  {\bibfield  {journal} {\bibinfo  {journal} {Phys. Rev. Lett.}\ }\textbf
  {\bibinfo {volume} {107}},\ \bibinfo {pages} {240501} (\bibinfo {year}
  {2011})}\BibitemShut {NoStop}%
\bibitem [{\citenamefont {Gr\"unhaupt}\ \emph {et~al.}(2018)\citenamefont
  {Gr\"unhaupt}, \citenamefont {Maleeva}, \citenamefont {Skacel}, \citenamefont
  {Calvo}, \citenamefont {Levy-Bertrand}, \citenamefont {Ustinov},
  \citenamefont {Rotzinger}, \citenamefont {Monfardini}, \citenamefont
  {Catelani},\ and\ \citenamefont
  {Pop}}]{QuasiparticlesInGranularAluminum2018}%
  \BibitemOpen
  \bibfield  {author} {\bibinfo {author} {\bibfnamefont {L.}~\bibnamefont
  {Gr\"unhaupt}}, \bibinfo {author} {\bibfnamefont {N.}~\bibnamefont
  {Maleeva}}, \bibinfo {author} {\bibfnamefont {S.~T.}\ \bibnamefont {Skacel}},
  \bibinfo {author} {\bibfnamefont {M.}~\bibnamefont {Calvo}}, \bibinfo
  {author} {\bibfnamefont {F.}~\bibnamefont {Levy-Bertrand}}, \bibinfo {author}
  {\bibfnamefont {A.~V.}\ \bibnamefont {Ustinov}}, \bibinfo {author}
  {\bibfnamefont {H.}~\bibnamefont {Rotzinger}}, \bibinfo {author}
  {\bibfnamefont {A.}~\bibnamefont {Monfardini}}, \bibinfo {author}
  {\bibfnamefont {G.}~\bibnamefont {Catelani}}, \ and\ \bibinfo {author}
  {\bibfnamefont {I.~M.}\ \bibnamefont {Pop}},\ }\href {\doibase
  10.1103/PhysRevLett.121.117001} {\bibfield  {journal} {\bibinfo  {journal}
  {Phys. Rev. Lett.}\ }\textbf {\bibinfo {volume} {121}},\ \bibinfo {pages}
  {117001} (\bibinfo {year} {2018})}\BibitemShut {NoStop}%
\bibitem [{\citenamefont {Maleeva}\ \emph {et~al.}(2018)\citenamefont
  {Maleeva}, \citenamefont {Grünhaupt}, \citenamefont {Klein}, \citenamefont
  {Levy-Bertrand}, \citenamefont {Dupre}, \citenamefont {Calvo}, \citenamefont
  {Valenti}, \citenamefont {Winkel}, \citenamefont {Friedrich}, \citenamefont
  {Wernsdorfer}, \citenamefont {Ustinov}, \citenamefont {Rotzinger},
  \citenamefont {Monfardini}, \citenamefont {Fistul},\ and\ \citenamefont
  {Pop}}]{GranularAlNonlinearity2018}%
  \BibitemOpen
  \bibfield  {author} {\bibinfo {author} {\bibfnamefont {N.}~\bibnamefont
  {Maleeva}}, \bibinfo {author} {\bibfnamefont {L.}~\bibnamefont {Grünhaupt}},
  \bibinfo {author} {\bibfnamefont {T.}~\bibnamefont {Klein}}, \bibinfo
  {author} {\bibfnamefont {F.}~\bibnamefont {Levy-Bertrand}}, \bibinfo {author}
  {\bibfnamefont {O.}~\bibnamefont {Dupre}}, \bibinfo {author} {\bibfnamefont
  {M.}~\bibnamefont {Calvo}}, \bibinfo {author} {\bibfnamefont
  {F.}~\bibnamefont {Valenti}}, \bibinfo {author} {\bibfnamefont
  {P.}~\bibnamefont {Winkel}}, \bibinfo {author} {\bibfnamefont
  {F.}~\bibnamefont {Friedrich}}, \bibinfo {author} {\bibfnamefont
  {W.}~\bibnamefont {Wernsdorfer}}, \bibinfo {author} {\bibfnamefont {A.~V.}\
  \bibnamefont {Ustinov}}, \bibinfo {author} {\bibfnamefont {H.}~\bibnamefont
  {Rotzinger}}, \bibinfo {author} {\bibfnamefont {A.}~\bibnamefont
  {Monfardini}}, \bibinfo {author} {\bibfnamefont {M.~V.}\ \bibnamefont
  {Fistul}}, \ and\ \bibinfo {author} {\bibfnamefont {I.~M.}\ \bibnamefont
  {Pop}},\ }\href {https://doi.org/10.1038/s41467-018-06386-9} {\bibfield
  {journal} {\bibinfo  {journal} {Nature Communications}\ }\textbf {\bibinfo
  {volume} {9}},\ \bibinfo {pages} {3889} (\bibinfo {year} {2018})}\BibitemShut
  {NoStop}%
\bibitem [{\citenamefont {Grünhaupt}\ \emph {et~al.}(2018)\citenamefont
  {Grünhaupt}, \citenamefont {Spiecker}, \citenamefont {Gusenkova},
  \citenamefont {Maleeva}, \citenamefont {Skacel}, \citenamefont {Takmakov},
  \citenamefont {Valenti}, \citenamefont {Winkel}, \citenamefont {Rotzinger},
  \citenamefont {Ustinov},\ and\ \citenamefont {Pop}}]{GralFluxonium2018}%
  \BibitemOpen
  \bibfield  {author} {\bibinfo {author} {\bibfnamefont {L.}~\bibnamefont
  {Grünhaupt}}, \bibinfo {author} {\bibfnamefont {M.}~\bibnamefont
  {Spiecker}}, \bibinfo {author} {\bibfnamefont {D.}~\bibnamefont {Gusenkova}},
  \bibinfo {author} {\bibfnamefont {N.}~\bibnamefont {Maleeva}}, \bibinfo
  {author} {\bibfnamefont {S.~T.}\ \bibnamefont {Skacel}}, \bibinfo {author}
  {\bibfnamefont {I.}~\bibnamefont {Takmakov}}, \bibinfo {author}
  {\bibfnamefont {F.}~\bibnamefont {Valenti}}, \bibinfo {author} {\bibfnamefont
  {P.}~\bibnamefont {Winkel}}, \bibinfo {author} {\bibfnamefont
  {H.}~\bibnamefont {Rotzinger}}, \bibinfo {author} {\bibfnamefont {A.~V.}\
  \bibnamefont {Ustinov}}, \ and\ \bibinfo {author} {\bibfnamefont {I.~M.}\
  \bibnamefont {Pop}},\ }\href@noop {} {\bibfield  {journal} {\bibinfo
  {journal} {arXiv:1809.10646}\ } (\bibinfo {year} {2018})}\BibitemShut
  {NoStop}%
\bibitem [{\citenamefont {Kou}\ \emph {et~al.}(2017)\citenamefont {Kou},
  \citenamefont {Smith}, \citenamefont {Vool}, \citenamefont {Brierley},
  \citenamefont {Meier}, \citenamefont {Frunzio}, \citenamefont {Girvin},
  \citenamefont {Glazman},\ and\ \citenamefont {Devoret}}]{3DFluxonium2017}%
  \BibitemOpen
  \bibfield  {author} {\bibinfo {author} {\bibfnamefont {A.}~\bibnamefont
  {Kou}}, \bibinfo {author} {\bibfnamefont {W.~C.}\ \bibnamefont {Smith}},
  \bibinfo {author} {\bibfnamefont {U.}~\bibnamefont {Vool}}, \bibinfo {author}
  {\bibfnamefont {R.~T.}\ \bibnamefont {Brierley}}, \bibinfo {author}
  {\bibfnamefont {H.}~\bibnamefont {Meier}}, \bibinfo {author} {\bibfnamefont
  {L.}~\bibnamefont {Frunzio}}, \bibinfo {author} {\bibfnamefont {S.~M.}\
  \bibnamefont {Girvin}}, \bibinfo {author} {\bibfnamefont {L.~I.}\
  \bibnamefont {Glazman}}, \ and\ \bibinfo {author} {\bibfnamefont {M.~H.}\
  \bibnamefont {Devoret}},\ }\href@noop {} {\bibfield  {journal} {\bibinfo
  {journal} {Phys. Rev. X}\ }\textbf {\bibinfo {volume} {7}},\ \bibinfo {pages}
  {031037} (\bibinfo {year} {2017})}\BibitemShut {NoStop}%
\end{thebibliography}
%

\end{document}